\documentclass[english,aps,twocolumn,superscriptaddress, aps, prapplied,longbibliography]{revtex4-2}
\usepackage[T1]{fontenc}
\usepackage[utf8]{inputenc}
\usepackage[active]{srcltx}
\usepackage{babel}
\usepackage{amsmath}
\usepackage{amssymb}
\usepackage{graphicx}
\usepackage[unicode=true,pdfusetitle,
 bookmarks=true,bookmarksnumbered=false,bookmarksopen=false,
 breaklinks=false,pdfborder={0 0 1},backref=false,colorlinks=false]
 {hyperref}

\makeatletter

\providecommand{\tabularnewline}{\\}

\usepackage[printonlyused]{acronym}

\AtBeginDocument{%
    \newwrite\bibnotes
    \def\bibnotesext{Notes.bib}
    \immediate\openout\bibnotes=\jobname\bibnotesext
    \immediate\write\bibnotes{@CONTROL{REVTEX41Control}}
    \immediate\write\bibnotes{@CONTROL{%
    apsrev41Control,author="08",editor="1",pages="1",title="0",year="1"}}
     \if@filesw
     \immediate\write\@auxout{\string\citation{apsrev41Control}}%
    \fi
}%

\makeatother

\begin{document}
\title{Optimizing the architecture for coherent beat note acquisition in
LISA}
\author{Philipp Euringer}
\email{philipp.euringer@airbus.com}

\affiliation{Airbus Space Systems, Airbus Defence and Space GmbH, Claude-Dornier-Straße,
88090 Immenstaad am Bodensee, Germany}
\author{Gerald Hechenblaikner}
\author{Alexander Sell}
\affiliation{Airbus Space Systems, Airbus Defence and Space GmbH, Claude-Dornier-Straße,
88090 Immenstaad am Bodensee, Germany}
\author{Francis Soualle}
\affiliation{Airbus Space Systems, Airbus Defence and Space GmbH, Willy-Messerschmitt-Straße
1, 82024 Taufkirchen, Germany}
\author{Walter Fichter}
\affiliation{University of Stuttgart, Pfaffenwaldring 27, 70569 Stuttgart, Germany}
\begin{abstract}
The laser interferometer space antenna (LISA) senses gravitational
waves by measuring distance fluctuations between three spacecraft
(SC). These measurements rely on precise tracking of a beat note phase
that is formed on a quadrant-photo-diode (QPD) at each SC by interference
of a local laser with a laser sent from a distant SC. The crucial
prerequisite of the phase tracking is a successful acquisition of
the beat note frequency. This article  aims to optimize the carrier-to-noise
density ratio (CNR) during this process, and to evaluate the resulting
probability of detection (PD). CNR is generally lowest during the
beat note acquisition process since pointing accuracy relies on coarse
acquisition techniques. Based on analytical models, we examine which
combinations of QPD segments for the signal read-out yield the highest
CNR, i.e., they are least susceptible to pointing errors. We find
from simulations that the highest CNR is ensured by taking the maximum
of a combination of two segments in vertical and horizontal direction.
For pointing errors (3$\sigma$) of 3.9 $\mu\text{rad}$ and 4.3 $\mu\text{rad}$
this yields an improvement of around 3.7 dB and 5.6 dB in CNR, respectively,
in comparison to a combination of all four segments. In addition,
the PD for various configurations of the baselined Fourier peak detection
is analyzed. Here we find that the PD is most sensitive to the CNR
compared to the design parameters of the acquisition scheme, in particular
the FFT length. Moreover, it is shown that aforementioned improvements
in CNR can lead to a significant enhancement of the PD. 
\end{abstract}
\maketitle

\section{Introduction\label{sec:Introduction}}

The laser interferometer space antenna (LISA) is a planned space-based
gravitational wave observatory that aims to detect gravitational waves
in a measurement bandwidth (BW) from 0.1 mHz to 1 Hz \citep{danzmann2011LISA}.
Detection requires a strain sensitivity of $10^{-21}$. This is achieved
using a constellation of three SC forming a triangle with a nominal
arm length of 2.5 million km. Between each pair of SC two one-way
optical links are established in opposing directions. At each of the
SC interference of the distant laser with a local one creates a beat
note, whose phase fluctuations are proportional to distance fluctuations
among the SC. To retrieve the gravitational wave signal the phase
of the beat note must be closely tracked via a phase-locked loop (PLL)
with a noise floor at 10 $\text{pm}/\sqrt{\text{Hz}}$.

However, prior to tracking the links must be acquired first. This
is performed in several steps. For the constellation acquisition,
a similar approach has been proposed for LISA \citep{Cirillo_2009}
and TAIJI \citep{Hechenblaikner2010abswave}. It comprises two major
phases: (1) spatial acquisition and (2) frequency acquisition. We
shall briefly discuss spatial acquisition for one optical link, where
SC1 is assumed to transmit a beam towards the receiving SC2. Owing
to the large initial uncertainty of the beam pointing (on the order
of tens of $\mu\text{rad}$) and the small divergence angle of the
laser beam (on the order of a few $\mu\text{rad}$), SC1 cannot directly
point its beam towards the presumed location of SC2, as the beam would
almost certainly miss it. Instead, it performs a search spiral in
the angular cone of uncertainty. At some point during the spiral scan,
the beam \textquotedblleft hits\textquotedblright{} SC2 and is detected
by its acquisition sensor (CAS) \citep{Hindman2004,Cirillo_2009}.
SC2 then re-orients itself towards the direction of the received beam
so that its own transmitted beam is detected by the scanning SC1 which
in turn re-orients itself in the direction of SC1. At this point,
the beam transmitted from each SC is imaged onto the CAS of the respective
opposite SC. Repeating this process for the two other link-pairs concludes
the spatial acquisition phase, see e.g. \citep{Hyde_2004,Cirillo_2009,Hechenblaikner2021PerfJitterAcq}.
The probability of success for spatial acquisition primarily depends
on the right balance between beam jitter, spiral pitch and beam divergence
angle \citep{Hechenblaikner2021PerfJitterAcq}. More refined models
also account for the jitter spectrum and its correlation function
\citep{Hechenblaikner2022SpecNoiseShape}, as well as the scan speed
and signal-to-noise ratio of the beam image on the detector \citep{Hindman2004,HECHENBLAIKNER2023InField,Hechenblaikner2023SearchBeam,Huang2023,Gao2020}.\\
Spatial acquisition is followed by frequency acquisition, also referred
to as \textquotedblleft coherent acquisition\textquotedblright , because
a coherent sensor, namely the long-arm interferometer with its QPD
sensors, is used to detect the received beam and subsequently lock
the PLL, thereby allowing nominal operation of the interferometers
\citep{EESA,GAO2023FullDemo}. In a first and most critical step,
which is the primary focus of this paper, the beat note between the
received beam and a local reference beam must be detected. This can
only be accomplished if the relative frequency between the two beams,
referred to as \textquotedblleft heterodyne frequency\textquotedblright ,
lies within the bandwidth of the phasemeter used for interferometric
detection. However, the relative frequency is only known to an accuracy
on the order of 100 MHz due to the uncertainties of ground-to-orbit
and of Doppler frequency shifts induced by the orbital dynamics. This
requires a scan of the local oscillator frequency over the entire
frequency uncertainty range. At the end of this scan the beat note
is detected by a maximum peak search as described in section \ref{subsec:Fourier-Acquisition}.
In a second step, which is less critical and represents no major
technical challenge, the PLL is locked after initializing its internal
numerically controlled oscillator (NCO) with the heterodyne frequency
found in the previous step. At this point, the phase can be read from
each of the 4 QPD segment channels and differential wavefront sensing
(DWS) \citep{Morrison1994,Hechenblaikner2010abswave} can be used
to stabilize the pointing with a precision on the order of a few nanorad
\citep{Cirillo_2009,GAO2023FullDemo}. Finally, a delay-locked-loop
(DLL) is activated to obtain inter-SC ranging information and extract
the incoming data stream from PRN code sequences modulated onto the
carrier frequency \citep{Esteban2009,heinzel2011auxiliary,EuringerTIM2023}.\\
Considerable efforts have been dedicated to the beat note acquisition
process itself \citep{BrauseThesis2018,EESA} and to various optimization
schemes \citep{Ales2015,Wang2023FFTCoG,Zhang2024MultiFreq}. However,
the read-out scheme at the QPD has received very little attention
in this context. This, however, is of particular importance since
it is well known that the CNR of the beat note strongly depends on
the angle of incidence of the interfering laser due to its linear
dependence on the heterodyne efficiency \citep{WANNER2012,Wanner2014,MahrdtPhD2014}.
During the beat note acquisition, the angle of incidence of the distant
laser is enlarged by pointing errors of receiving and transmitting
SC. During nominal operation, these pointing errors are minimized
via DWS, which, however, is not applicable during the beat note acquisition
as it requires coherent tracking. Consequently, we are aiming for
a read-out scheme that is least susceptible to pointing errors and
thus maximizes the CNR. This is the central point of the presented
analysis. \\
Based on a careful analysis of the heterodyne efficiency on the pointing
error of the receiving SC, we show that the maximum heterodyne efficiency
of two single segments surpasses the heterodyne efficiency of two
segments combined along any axis of the QPD. Moreover the heterodyne
efficiency is always most susceptible to pointing errors when all
four segments are combined. For these three configurations, the sensitivity
of the CNR to pointing errors of transmitting and receiving SC is
analyzed. We show that a suitable selection of the configuration
can increase the minimum CNR during the beat note acquisition by
up to 5.6 dB. In addition, we provide an evaluation of the current
beat note acquisition scheme and show that the improvement in CNR
can result in a significant enhancement of the PD.

The paper is divided in two parts. Section \ref{sec:CNR-of-Acquisition}
is dedicated to the analysis of the CNR. In section \ref{subsec:Generic-Model}
a general model for the CNR analysis is introduced. To this end,
reasonable assumptions are taken into account to allow separate evaluation
of parameters affected by pointing errors of the transmitting and
the receiving SC. First, pointing errors of the transmitting SC in
terms of the received power are analyzed in section \ref{subsec:Angle-dependence-of-power}.
Pointing errors of the receiving SC are analyzed in section \ref{subsec:Angle-dependence-of}
based on the heterodyne efficiency, which naturally leads to the definition
of three different read-out configurations. After a discussion of
relevant noise contributions in section \ref{subsec:Noise-Contributions},
the CNR of the three configurations in terms of pointing errors of
receiving and transmitting SC is analyzed in section \ref{subsec:Angle-dependence-CNR}.\\
Section \ref{sec:Application:-Fourier-Acquisition} analyses the influence
of the CNR on the beat note acquisition scheme. The basic principle
and model assumption on the beat note acquisition scheme are described
in section \ref{subsec:Fourier-Acquisition}. In section \ref{subsec:Probability-of-detection}
an analysis on the PD is performed. A discussion of the analysis and
the implementation of the configurations follows in section \ref{subsec:Discussion:-Applicability-of}.

\section{CNR during beat note acquisition\label{sec:CNR-of-Acquisition}}

\subsection{Generic model\label{subsec:Generic-Model}}

\begin{table}
\centering{}\small\caption{Parameter definition\label{tab:Parameter-definition}}
\begin{tabular}{ccc}
\hline 
\hline Symbol & Unit & Description\tabularnewline
\hline 
$a_{\text{lo,1f}}$,$a_{\text{s,1f}}$ & 1/$\sqrt{\text{Hz}}$ & 1f-RIN of local oscillator/signal beam\tabularnewline
$C$ & $\text{V}^{2}$ & carrier power\tabularnewline
CNR & Hz & carrier-to-noise density ratio\tabularnewline
$\text{d}S$ & $\text{m}^{2}$ & surface element\tabularnewline
$d_{\text{sc}}$ & m & inter-SC distance\tabularnewline
$E_{\text{lo}}$,$E_{\text{s}}$ & V/$\text{m}^{2}$ & electric field of local/signal beam\tabularnewline
$E_{\text{lo}}^{0}$,$E_{\text{s}}^{0}$ & V/$\text{m}^{2}$ & amplitude of local/signal beam\tabularnewline
$F_{\text{CNR,lim}}$ & - & worst-case CNR\tabularnewline
$f_{\text{s}}$ & Hz & sampling rate\tabularnewline
$f_{\text{res}}$ & Hz & frequency resolution of FFT\tabularnewline
$k_{\text{lo}}$,$k_{\text{s}}$ & rad/m & wave number of local/signal beam\tabularnewline
$K$,$K_{\text{tot}}$ & - & number of bins/total bins per acquisition\tabularnewline
$L$ & - & number of samples per FFT\tabularnewline
$M$ & - & number of FFT intervals\tabularnewline
$M_{2}$ & - & total magnification on QPD\tabularnewline
$m_{\text{sb}}$ & rad & sideband modulation index\tabularnewline
$N_{0}$ & $\text{V}^{2}/\text{Hz}$ & one-sided Noise PSD\tabularnewline
$N_{\text{QPD}}$ & - & number of combined QPDs\tabularnewline
$N_{\text{s}}$ & - & number of segments combined per QPD\tabularnewline
$n_{\text{scans}}$ & - & number of scans\tabularnewline
$P_{\text{lo}}$,$P_{\text{s}}$ & W & power of local/signal beam at QPD\tabularnewline
$P_{\text{s,n}}$ & W & nominal received power\tabularnewline
$P$ & W & power at the QPD\tabularnewline
$P_{\text{D}}$,$P_{\text{D,n}}$ & - & PD with/without sidebands\tabularnewline
$P_{\text{s}}$ & W & power of signal beam at QPD\tabularnewline
$R_{\text{min}}$ & m & $\min\left(R_{\text{tel}}/M_{2},R_{\text{QPD}},R_{\text{i}}\right)$\tabularnewline
$R_{\text{i}}$,$R_{\text{tel}}$ & m & internal/external pupil radius\tabularnewline
$R_{\text{QPD}}$ & m & radius of detector area\tabularnewline
$S_{\text{bin}}$ & $\text{V}^{2}$ & bin power\tabularnewline
$s_{\text{I,seg,el}}$ & A & current noise density of QPD electronics\tabularnewline
$W_{0}$ & m & 1/$e^{2}$ radius of local beam on QPD\tabularnewline
$x,y,z$ & m & axes of the optical reference frame\tabularnewline
$x_{\text{s}},y_{\text{s}},z_{\text{s}}$ & m & axes of the signal reference frame\tabularnewline
$z_{0}$ & m & Rayleigh range\tabularnewline
$\alpha$,$\beta$ & rad & angle in/out of constellation plane at QPD\tabularnewline
$\alpha_{\text{rx}}$,$\beta_{\text{rx}}$ & rad & angle in/out of constellation plane at Rx SC\tabularnewline
$\alpha_{\text{tx}}$,$\beta_{\text{tx}}$ & rad & angle in/out of constellation plane at Tx SC\tabularnewline
$\gamma_{\text{max}}$ & rad & magnitude of $\alpha_{\text{rx}}$ and $\beta_{\text{rx}}$\tabularnewline
$\gamma_{\text{tx}}$ & rad & magnitude of $\alpha_{\text{tx}}$ and $\beta_{\text{tx}}$\tabularnewline
$\eta_{\text{carrier}}$ & - & portion of optical power in main carrier\tabularnewline
$\eta_{\text{het}}$ & - & heterodyne efficiency\tabularnewline
$\eta_{\text{resp}}$ & A/W & responsitvity of QPD\tabularnewline
$\eta_{\text{wfe}}$ & - & wave front error\tabularnewline
$\lambda_{\text{s}}$ & m & wavelength of signal beam\tabularnewline
$\delta$ & - & non-centrality parameter\tabularnewline
$\delta_{\text{main}}$,$\delta_{\text{sb}}$ & - & $\delta$ for main/sideband beat note\tabularnewline
$\sigma_{\text{bin}}^{2}$ & $\text{V}^{2}$ & variance of noise bin \tabularnewline
$\phi$ & rad & radial angle\tabularnewline
$\varphi_{\text{t}}$,$\varphi_{\text{i}}$ & rad & total/interferometric phase of beat note\tabularnewline
$\omega_{\text{lo}}$,$\omega_{\text{s}}$ & rad/s & angular frequency of local/signal beam\renewcommand{\tabularnewline}{\\ \hline}\tabularnewline
\hline 
\end{tabular}\renewcommand{\tabularnewline}{\\}
\end{table}
 \normalsize Spatial acquisition of LISA has been completed once
each of the six lasers is detected at the center of the CAS of the
adjacent SC. The pointing accuracy is represented by random variables
$\alpha$ and $\beta$ for the angle in constellation plane and out
of constellation plane, respectively. For completeness, these and
all subsequently defined parameters are listed in Tab. \ref{tab:Parameter-definition}.
If not otherwise specified these parameters are assigned with values
listed in \citep{LISAPRFMODEL}, thus providing a LISA-representative
evaluation.
\begin{figure}
\begin{centering}
\includegraphics[width=\linewidth]{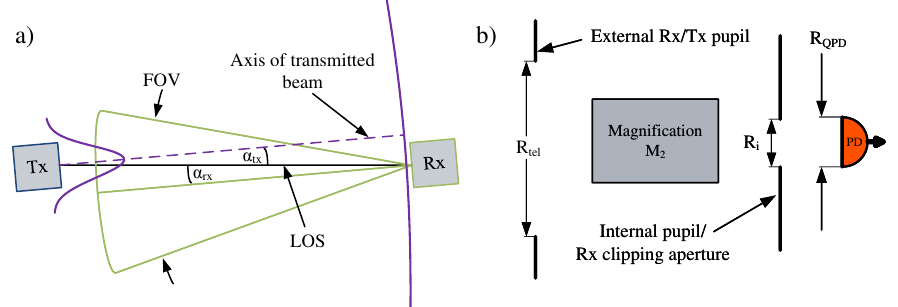}
\par\end{centering}
\centering{}\caption{In a) transmitting ('Tx') and receiving ('Rx') SC are depicted by
gray boxes. The field of view (FOV) of the receiving SC is depicted
by the green cone. The deviation of the FOV center line to the line
of sight (LOS) is indicated by the angle $\alpha_{\text{rx}}$. Purple
lines indicate the intensity profile of the distant laser close to
the transmitting SC and at the reception of the receiving SC. The
deviation of the axis of the transmited beam to the LOS is indicated
by the angle $\alpha_{\text{tx}}$. In b) the apertures in the receive
and transmit path are identified. The photodiode is indicated with
'PD'. \label{fig:Idenfication-of-apertures}}
\end{figure}
\\
In general, each laser link is affected by the pointing error of the
transmitting SC (denoted via subscript 'tx') and the receiving SC
(denoted via subscript 'rx'), leading to four angles $\alpha_{\text{tx}},\beta_{\text{tx}},\alpha_{\text{rx}}$
and $\beta_{\text{rx}}$, for each laser link. The transmitting angles
define which portion of the light cone is present at the receiving
SC, whereas both, receiving and transmitting angles, define the angle
of incidence at the receiving SC, see Fig. \ref{fig:Idenfication-of-apertures}
a). The 3$\sigma$ values of $\alpha_{\text{tx}},\beta_{\text{tx}},\alpha_{\text{rx}}$
and $\beta_{\text{rx}}$ are assumed to be identical and are assigned
with the values listed in Tab. \ref{tab:Pointing-accuracy.}. The
CNR of 61 dB-Hz, listed in \citep{Bachman2017}, as the lower limit
for DWS, has been used to derive a value of 3.9 $\mu\text{rad}$ (3$\sigma$)
for the tolerable angular uncertainty. Option 2 assumes a slightly
more robust DWS operation with respect to option 1.
\begin{table}
\centering{}\caption{Pointing error of angles $\alpha_{\text{tx}},\beta_{\text{tx}},\alpha_{\text{rx}}$
and $\beta_{\text{rx}}$\label{tab:Pointing-accuracy.}}
\begin{tabular}{cc}
\hline 
\hline Option & 3$\sigma$\tabularnewline
\hline 
1 & 3.9 $\mu\text{rad}$\tabularnewline
\hline 
2 & 4.3 $\mu\text{rad}$\renewcommand{\tabularnewline}{\\ \hline}\tabularnewline
\end{tabular}\renewcommand{\tabularnewline}{\\}
\end{table}
 \\
At the QPD, the beam of the distant SC interferes with a beam
of the local SC. We will denote them as signal beam and local oscillator
beam, respectively. The signal beam enters the receiving SC through
the telescope with radius $R_{\text{tel}}$, see Fig. \ref{fig:Idenfication-of-apertures}
b). Since transmitting and receiving SC are separated by $\sim2.5$
million km, the wavefront curvature is negligible and the intensity
distribution is flat so that the received beam is well described by
a plane wave over the limiting aperture of the telescope. Consequently,
in this case, the complex electric field of the signal beam (indicated
via subscript 's'), in its local coordinate system, denoted as signal
reference frame (SRF), reads as 
\begin{equation}
E_{\text{s}}=E_{\text{s}}^{0}(\alpha_{\text{tx}},\beta_{\text{tx}})e^{\text{i}k_{\text{s}}z_{\text{s}}}e^{-\text{i}\omega_{\text{s}}t}.\label{eq:def-signal-beam}
\end{equation}
Hereby, $\omega_{\text{s}}$ denotes the angular frequency and $k_{\text{s}}$
($\approx2\pi/1064$ rad $\text{nm}^{-1}$) the wave vector of the
signal beam. We note that based on the beam characteristics, the
transmitted beam angles $\alpha_{\text{tx}}$ and $\beta_{\text{tx}}$
only affect the amplitude $E_{\text{s}}^{0}$.\\
The local oscillator beam incident at the QPD is modeled as a Gaussian
beam. Its complex representation in paraxial approximation reads as
\citep{Saleh_2019}
\begin{align}
E_{\text{lo}} & =E_{\text{lo}}^{0}\exp\left(-\dfrac{x^{2}+y^{2}}{W^{2}(z)}\right)\cdot\nonumber \\
 & \;\;\cdot\exp\left(\text{i}k_{\text{lo}}z+\text{i}k_{\text{lo}}\dfrac{x^{2}+y^{2}}{2R(z)}-\text{i}\zeta(z)-\text{i}\omega_{\text{lo}}t\right),\label{eq:def_local_beam}
\end{align}
with 
\begin{align*}
W(z) & =W_{0}\sqrt{1+(z/z_{0})^{2}},\\
R(z) & =z\left[1+(z_{0}/z)^{2}\right],\\
\zeta(z) & =\text{atan}\left(z/z_{0}\right),\\
W_{0} & =\sqrt{\lambda z_{0}/\pi}.
\end{align*}
The parameter $\omega_{\text{lo}}$ denotes the angular frequency
and $k_{\text{lo}}$ ($\approx2\pi/1064$ rad $\text{nm}^{-1}$) the
wave vector of the local beam. Amplitude and Rayleigh range are denoted
as $E_{\text{lo}}^{0}$ and $z_{0}$, respectively. The local oscillator
beam is expressed in the optical reference frame (ORF) with associated
coordinates ($x,y,z$). 
\begin{figure}
\begin{centering}
\includegraphics[width=.7\linewidth]{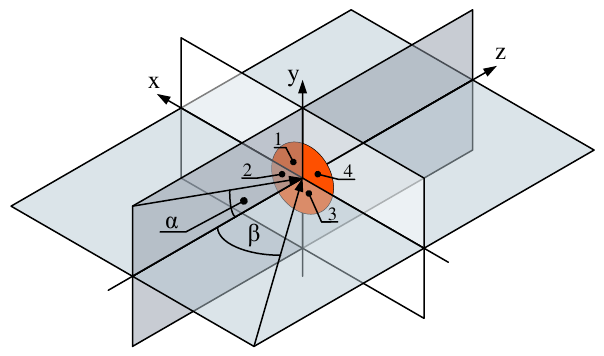}
\par\end{centering}
\centering{}\caption{Definition of the ORF and angles $\alpha$ and $\beta$. The QPD
is illustrated by an orange circle, where the segments are indicated.\label{fig:Definition-of-COS,}}
\end{figure}
For the ongoing analysis we assume that the QPD is perfectly aligned
with the optical axis of the receiving SC, i.e. with the ORF. We
set the origin of the ORF at the center of the QPD as indicated in
Fig. \ref{fig:Definition-of-COS,}. Moreover, we assume that the local
oscillator beam is perfectly aligned and focused on the QPD, as evident
from Eq. \ref{eq:def_local_beam}.\\
To express Eq. \ref{eq:def-signal-beam} in the ORF, no translation
along the $x$ and $y$-axis needs to be considered thanks to its
beam characteristics. Moreover, the rotation around the $z$-axis
is ignored due to axial symmetry. Consequently, \small
\begin{align}
{\displaystyle \begin{aligned}\begin{bmatrix}\begin{array}{c}
x_{\text{s}}\\
y_{\text{s}}\\
z_{\text{s}}
\end{array}\end{bmatrix} & =\begin{bmatrix}\cos\beta & 0 & \sin\beta\\
0 & 1 & 0\\
-\sin\beta & 0 & \cos\beta
\end{bmatrix}\begin{bmatrix}1 & 0 & 0\\
0 & \cos\alpha & -\sin\alpha\\
0 & \sin\alpha & \cos\alpha
\end{bmatrix}\begin{bmatrix}\begin{array}{c}
x\\
y\\
z
\end{array}\end{bmatrix}+\begin{bmatrix}\begin{array}{c}
0\\
0\\
d_{\text{sc}}
\end{array}\end{bmatrix}\end{aligned}
}\label{eq:def_trans_cos}
\end{align}
\normalsize relates the ORF to the SRF. The angles $\alpha$ and
$\beta$ are indicated in Fig. \ref{fig:Definition-of-COS,} and represent
the angle of the wave vector of the incident plane wave projected
onto the $y$-$z$ plane and $x$-$z$ plane, respectively. These
angles are expressed relative to the inertial (on-sky) coordinate
system by division with the magnification $M_{2}$, which establishes
the relation $\alpha_{\text{rx}}=\alpha/M_{2}$ and $\beta_{\text{rx}}=\beta/M_{2}$.
The distance between the SC is denoted as $d_{\text{sc}}$. \\
Inserting Eq. \ref{eq:def_trans_cos} in Eq. \ref{eq:def-signal-beam},
allows us to write the signal beam as a function of the angles $\alpha_{\text{tx}},\beta_{\text{tx}},\alpha_{\text{rx}}$
and $\beta_{\text{rx}}$. To express the signal beam at the reception
of the QPD the limiting aperture needs to be taken into account, which
is incorporated as follows
\begin{align}
E_{\text{s}}(\alpha_{\text{tx}},\beta_{\text{tx}},\alpha_{\text{rx}},\beta_{\text{rx}}) & =\nonumber \\
\theta(R_{\text{min}}-\sqrt{x^{2}+y^{2}}) & E_{\text{s}}^{0}(\alpha_{\text{tx}},\beta_{\text{tx}})e^{-\text{i}\omega_{\text{s}}t}\cdot\nonumber \\
\cdot\exp(\text{i}k_{\text{s}}(-x\sin(M_{2} & \beta_{\text{rx}})+y\sin(M_{2}\alpha_{\text{rx}})\cos(M_{2}\beta_{\text{rx}})))\cdot\nonumber \\
\cdot\exp(\text{i}k_{\text{s}}(z\cos(M_{2} & \alpha_{\text{rx}})\cos(M_{2}\beta_{\text{rx}})+d_{\text{sc}})).\label{eq:redef_sig_beam}
\end{align}
with 
\[
R_{\text{min}}:=\min\left(R_{\text{tel}}/M_{2},R_{\text{QPD}},R_{\text{i}}\right).
\]
Here, the Heaviside step function $\theta$ takes into account the
clipping of the signal beam by the receiving optic. According to Fig.
\ref{fig:Idenfication-of-apertures} b), the beam diameter is either
limited by the ratio of the telescope aperture with radius $R_{\text{tel}}$
and the magnification $M_{2}$, the internal pupil with radius $R_{\text{i}}$
or the detector area with radius $R_{\text{QPD}}$. Using Eq. \ref{eq:redef_sig_beam}
and Eq. \ref{eq:def_local_beam}, we can then express the optical
power $P$ at the QPD resulting from the electrical field of signal
and local oscillator beam
\begin{align}
P & =\dfrac{1}{2}c\epsilon_{0}\iint\left|E_{\text{s}}(\alpha_{\text{tx}},\beta_{\text{tx}},\alpha_{\text{rx}},\beta_{\text{rx}})+E_{\text{lo}}\right|^{2}\,\text{d}S\nonumber \\
 & =\dfrac{N_{\text{s}}}{4}P_{\text{s}}(\alpha_{\text{tx}},\beta_{\text{tx}})+\dfrac{N_{\text{s}}}{4}P_{\text{lo}}\nonumber \\
 & +c\epsilon_{0}\Re\left(\iint E_{\text{s}}^{\ast}(\alpha_{\text{tx}},\beta_{\text{tx}},\alpha_{\text{rx}},\beta_{\text{rx}})E_{\text{lo}}\,\text{d}S\right).\label{eq:P_het_general}
\end{align}
Hereby, $c$ denotes the speed of light and $\epsilon_{0}$ the vacuum
permittivity. $E_{\text{s}}^{\ast}$ represents the complex conjugate
of $E_{\text{s}}$. Note that the surface element $\text{d}S=\text{d}x\text{d}y$
is integrated over an area (specified by integral boundaries) that
corresponds to either a single segment or combinations thereof.\\
As will be seen in section \ref{subsec:Angle-dependence-of}, for
the ongoing analysis, of most relevance are single segments, the combination
of two adjacent segments, and the combination of all four segments.
Due to symmetry, segment 1 (denoted with subscript 'seg 1') yields
the same CNR as segment 3 and similarly, segment 2 (denoted with subscript
'seg 2') equals segment 4. In addition, the vertical combination of
segment 1 and segment 2 (denoted with subscript 'ver') yields the
same CNR as the combination of segment 3 and segment 4. Similarly,
the horizontal combination of segment 1 and segment 4 (denoted with
subscript 'hor') is identical to the combination of segment 2 and
segment 3. Finally, taking into account the combination of all four
segments (denoted with subscript 'all') leaves five different segment
and segment combinations. These are specified via their integral
boundaries in Tab. \ref{tab:QPD-configurations.}, where columns indicated
with $x^{\pm}$ and $y^{\pm}$ denote the upper ($+$) and lower ($-$)
integral bounds. 
\begin{table}
\centering{}\caption{Segment combinations\label{tab:QPD-configurations.}}
\begin{tabular}{ccccc}
\hline 
\hline Label & $x^{-}$ & $x^{+}$ & $y^{-}$ & $y^{+}$\tabularnewline
\hline 
seg 1 & 0 & $\sqrt{R_{\text{min}}^{2}-y^{2}}$ & 0 & $R_{\text{min}}$\tabularnewline
seg 2 & 0 & $\sqrt{R_{\text{min}}^{2}-y^{2}}$ & $-R_{\text{min}}$ & 0\tabularnewline
ver & 0 & $\sqrt{R_{\text{min}}^{2}-y^{2}}$ & $-R_{\text{min}}$ & $R_{\text{min}}$\tabularnewline
hor & $-\sqrt{R_{\text{min}}^{2}-y^{2}}$ & $\sqrt{R_{\text{min}}^{2}-y^{2}}$ & 0 & $R_{\text{min}}$\tabularnewline
all & $-\sqrt{R_{\text{min}}^{2}-y^{2}}$ & $\sqrt{R_{\text{min}}^{2}-y^{2}}$ & $-R_{\text{min}}$ & $R_{\text{min}}$\renewcommand{\tabularnewline}{\\ \hline}\tabularnewline
\hline 
\end{tabular}\renewcommand{\tabularnewline}{\\}
\end{table}
In the following, the label will be attached as subscript to parameters
and integral bounds to indicate the respective segment (combination).
\\
In the second line of Eq. \ref{eq:P_het_general} we considered that
the optical power of the individual beams $P_{\text{s}/\text{lo}}:=\tfrac{1}{2}c\epsilon_{0}\underset{\text{all}}{\iint}\left|E_{\text{s}/\text{lo}}\right|^{2}\,\text{d}S$
is rotationally symmetric. Therefore, the power only depends on the
number $N_{\text{s}}$ of QPD segments considered for the readout
but not on the explicit segments. Only the last term, which is formed
by the real part $\Re$ of the interfering signal, explicitly depends
on the choice of the segments. This oscillating term, commonly known
as beat note, is usually expressed in terms of the heterodyne efficiency
$\eta_{\text{het}}$ defined according to \citep{Heinzel2020}
\begin{equation}
\sqrt{\eta_{\text{het}}}e^{-\text{i}\varphi_{\text{t}}}:=\dfrac{\iint E_{\text{s}}^{\ast}(\alpha_{\text{tx}},\beta_{\text{tx}},\alpha_{\text{rx}},\beta_{\text{rx}})E_{\text{lo}}\,\text{d}S}{\sqrt{\iint\left|E_{\text{s}}\right|^{2}\,\text{d}S\,\iint\left|E_{\text{lo}}\right|^{2}\,\text{d}S}}.\label{eq:intro_het_eff}
\end{equation}
The phase $\varphi_{\text{t}}=(\omega_{\text{s}}-\omega_{\text{lo}})t+\varphi_{\text{i}}$
constitutes the total phase that is formed by the beat note of distant
and local laser as well as the interferometric phase $\varphi_{\text{i}}$.
Consequently, only the last term in Eq. \ref{eq:P_het_general} represents
the signal of interest. As shown in appendix \ref{sec:Appendix-A-Heterodyne}
(Eq. \ref{eq:eta_het_intro}), inserting the explicit expressions
of the signal beam (Eq. \ref{eq:redef_sig_beam}) and the local oscillator
beam (Eq. \ref{eq:def_local_beam}) in Eq. \ref{eq:intro_het_eff},
the heterodyne efficiency can be expressed as
\begin{equation}
\eta_{\text{het}}=|\dfrac{2}{N_{\text{s}}}\dfrac{N(R_{\text{QPD}}/W_{0})}{W_{0}R_{\text{min}}\pi}E_{\text{rl}}(M_{2}\alpha_{\text{rx}},M_{2}\beta_{\text{rx}})|^{2}\label{eq:def_het_eff}
\end{equation}
with
\begin{align}
E_{\text{rl}}(\alpha,\beta) & :=\iint\exp\left(-\left(y^{2}/W_{0}^{2}+\text{i}k_{\text{s}}y\sin(\alpha)\cos(\beta)\right)\right)\cdot\nonumber \\
 & \cdot\exp\left(-\left(x^{2}/W_{0}^{2}-\text{i}k_{\text{s}}x\sin(\beta)\right)\right)\text{d}\tilde{S},\label{eq:eta_het_integ}\\
N(x) & :=2^{3/2}\left(1-e^{-2x^{2}}\right)^{-1/2}.\nonumber 
\end{align}
Hereby, $\text{d}\tilde{S}$ indicates that, in contrast to $\text{d}S$,
integration is only performed over the area that is not clipped by
the aperture with radius $R_{\text{min}}$. We note that Eq. \ref{eq:def_het_eff}
only depends on the angles of the received beam $\alpha_{\text{rx}}$
and $\beta_{\text{rx}}$ but not on $\alpha_{\text{tx}}$ and $\beta_{\text{tx}}$.
In a similar way this can also be shown for the phase $\varphi_{\text{t}}$.
Finally, using the relation, given by Eq. \ref{eq:intro_het_eff},
between the heterodyne efficiency of Eq. \ref{eq:def_het_eff} and
the interference term of Eq. \ref{eq:P_het_general}, we can rewrite
Eq. \ref{eq:P_het_general} after some re-arrangements and using Euler's
formula as follows:
\begin{align}
P & =\dfrac{N_{\text{s}}}{4}\left(P_{\text{s}}(\alpha_{\text{tx}},\beta_{\text{tx}})+P_{\text{lo}}\right)\nonumber \\
 & +\dfrac{N_{\text{s}}}{4}2\sqrt{P_{\text{s}}(\alpha_{\text{tx}},\beta_{\text{tx}})P_{\text{lo}}}\sqrt{\eta_{\text{het}}(M_{2}\alpha_{\text{rx}},M_{2}\beta_{\text{rx}})}\cdot\nonumber \\
 & \cdot\cos((\omega_{\text{s}}-\omega_{\text{lo}})t+\varphi_{\text{i}}(M_{2}\alpha_{\text{rx}},M_{2}\beta_{\text{rx}})).\label{eq:redef_P_het_general}
\end{align}
A similar expression has also been found in \citep{MahrdtPhD2014}.
The last term in Eq. \ref{eq:redef_P_het_general} represents the
beat note signal between signal and local oscillator beam. This signal
is then converted to a photovoltage signal using a transimpedance
amplifier. Based on Eq. \ref{eq:redef_P_het_general} it is then straightforward
to show that the beat note power $C$ of this photovoltage signal
is given by \small
\begin{align}
C= & 2\left(\dfrac{\eta_{\text{carrier}}\eta_{\text{resp}}N_{\text{s}}N_{\text{QPD}}}{16}\right)^{2}\cdot\nonumber \\
 & \cdot P_{\text{s}}(\alpha_{\text{tx}},\beta_{\text{tx}})P_{\text{lo}}\eta_{\text{wfe}}\eta_{\text{het}}(M_{2}\alpha_{\text{rx}},M_{2}\beta_{\text{rx}}).\label{eq:carrier_power}
\end{align}
\normalsize Equation \ref{eq:carrier_power} accounts for the efficiency
of the QPD via the responsitvity $\eta_{\text{resp}}$ and the fact
that the signal may combined from $N_{\text{QPD}}$ QPDs. In addition,
it is considered that parts of the carrier are modulated and therefore
only $\eta_{\text{carrier}}$ of the beat note can be used for the
phase read-out \citep{heinzel2011auxiliary,Delgado2012}. Moreover,
$\eta_{\text{wfe}}$ is an additional constant factor accounting for
wavefront errors. A numerical analysis on the wavefront errors including
a LISA representative value has been provided in \citep{LISATTL_UKOB}.
Importantly, Eq. \ref{eq:carrier_power} allows a separation of the
angle pairs $\alpha_{\text{rx}}$ and $\beta_{\text{rx}}$, which
affect $\eta_{\text{het}}$, from the angle pairs $\alpha_{\text{tx}}$
and $\beta_{\text{tx}}$, which determine the received power of the
signal beam $P_{\text{s}}$. Based on this separation, the sensitivity
of the received power on the pointing error of the transmitting SC
and the sensitivity of heterodyne efficiency on the pointing error
of the receiving SC will be examined in the subsequent sections.

\subsection{Angle dependence of received power\label{subsec:Angle-dependence-of-power}}

Each laser serves as local oscillator for the long-arm interferometer
of the local OB and as signal beam for long-arm interferometer of
the adjacent SC. The signal beam leaves the remote SC through the
optics shown in Fig. \ref{fig:Idenfication-of-apertures} b). Consequently,
the complex electric field of the signal beam in its local coordinate
system equals Eq. \ref{eq:def_local_beam} by changing the beam waist
and the laser frequency, which must be different for the interfering
beams. In particular, we make the following substitutions $W_{0}\rightarrow W_{0}M_{2}$,
$k_{\text{lo}}\rightarrow k_{\text{s}}$ and $\omega_{\text{lo}}\rightarrow\omega_{\text{s}}$.
The power $P_{\text{s}}(\alpha_{\text{tx}},\beta_{\text{tx}})$ of
the signal beam received at the local SC can be estimated using the
Fraunhofer approximation. This approximation is well suited due to
the large distance between the SC. We note that dedicated models
for wave front errors or effects due to optical aberration are not
considered in this model. Instead, these effects will be considered
via constant loss factors, which account for a nominal performance
degradation of these effects. Following the derivation of Appendix
\ref{sec:Appendix-C-Received}, the power $P_{\text{s}}$ received
through an aperture with radius $R_{\text{tel}}$ at a distance $d_{\text{sc}}$
is then given by
\begin{align}
P_{\text{s}}(\gamma_{\text{tx}}) & =2P_{\text{s,n}}\,\left(\dfrac{k_{\text{s}}R_{\text{tel}}^{2}}{d_{\text{sc}}}\right)^{2}F^{2}\left(\dfrac{R_{\text{tel}}}{W_{0}M_{2}},k_{\text{s}}R_{\text{tel}}\gamma_{\text{tx}}\right)\label{eq:power_s}
\end{align}
with 
\[
F(r,\kappa):=r\int_{0}^{1}\rho e^{-r^{2}\rho^{2}}J_{0}(\kappa\rho)\,\text{d}\rho
\]
where, $J_{0}$ denotes the zero order Bessel function. We note that
the received power only depends on the magnitude of the combined angle
$\gamma_{\text{tx}}=\sqrt{\alpha_{\text{tx}}^{2}+\beta_{\text{tx}}^{2}}$.
The parameter $P_{\text{s,n}}$ denotes the nominal power and accounts
for losses in the on-board transmit and receive path. 
\begin{figure}
\begin{centering}
\includegraphics[width=0.9\linewidth]{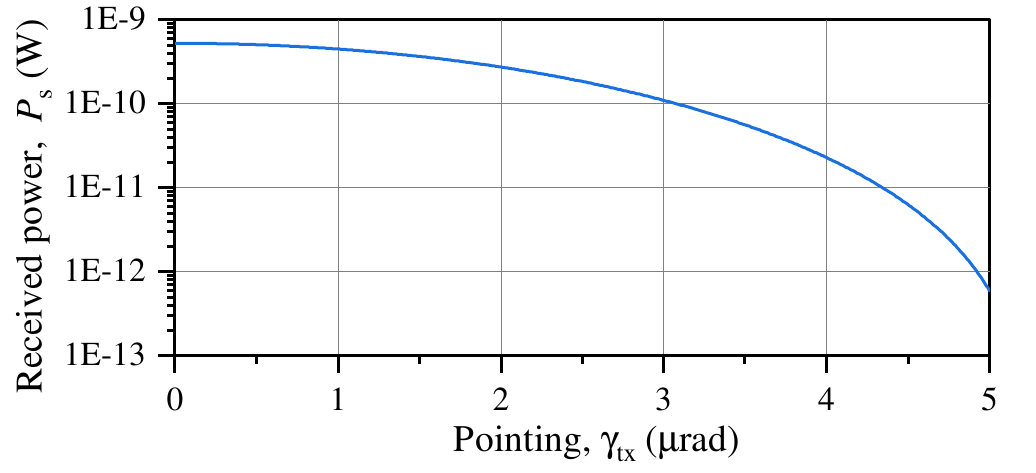}
\par\end{centering}
\centering{}\caption{Received Power under variation of the pointing angle $\gamma_{\text{tx}}$.\label{fig:Received-Power-under}}
\end{figure}
Figure \ref{fig:Received-Power-under} depicts the power $P_{\text{s}}$
under variation of the angle $\gamma_{\text{tx}}$. A minimum of the
received power is observed at around 5.3 $\mu\text{rad}$.

\subsection{Angle dependence of the heterodyne efficiency\label{subsec:Angle-dependence-of}}

According to Eq. \ref{eq:def_het_eff} and Eq. \ref{eq:eta_het_integ},
the heterodyne efficiency explicitly depends on the segments considered
for the read-out of the beat note. In the following we want to analyze
the sensitivity of the heterodyne efficiency on the receiving angles
$\alpha_{\text{rx}}$ and $\beta_{\text{rx}}$ taking into account
different QPD segments for the read-out. To this end, we will omit
effects resulting from the finite slit size separating the QPD segments.
We consider the five different configurations listed in Tab. \ref{tab:QPD-configurations.}. 

\begin{figure}
\begin{centering}
\includegraphics[width=\linewidth]{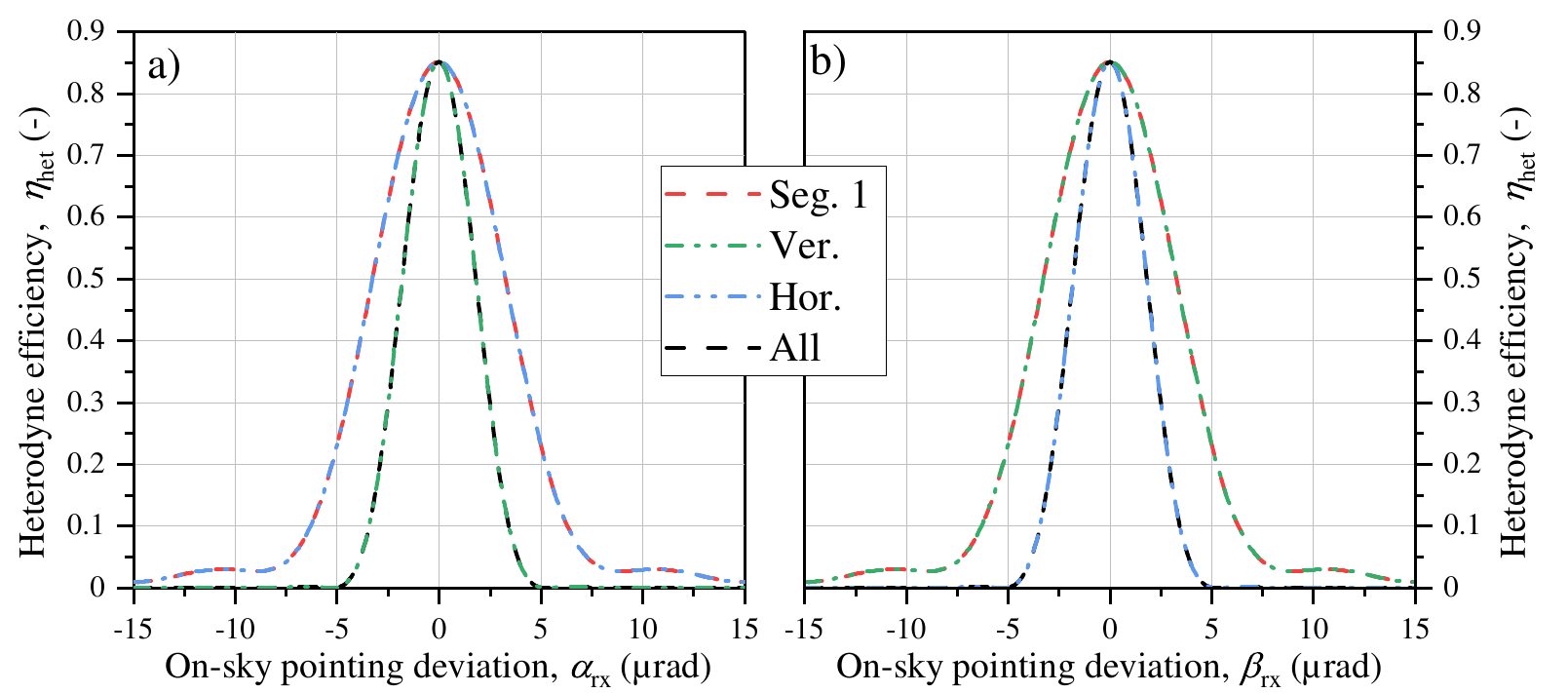}
\par\end{centering}
\centering{}\caption{The heterodyne efficiency $\eta_{\text{het}}$ as a function of angles
$\alpha_{\text{rx}}$ and $\beta_{\text{rx}}$ is shown in a) and
b), respectively. Red lines show the heterodyne efficiency for a single
segment, specifically for segment 1. The green curves depict the heterodyne
efficiency once the read-out of segment 1 and segment 2 are combined,
while the light blue curves combine segment 1 with segment 4. Finally,
the resulting heterodyne efficiency once all four segments are combined
is depicted by the black lines. \label{fig:The-heterodyne-efficiency}}
\end{figure}
\begin{figure}
\begin{centering}
\includegraphics[width=0.9\linewidth]{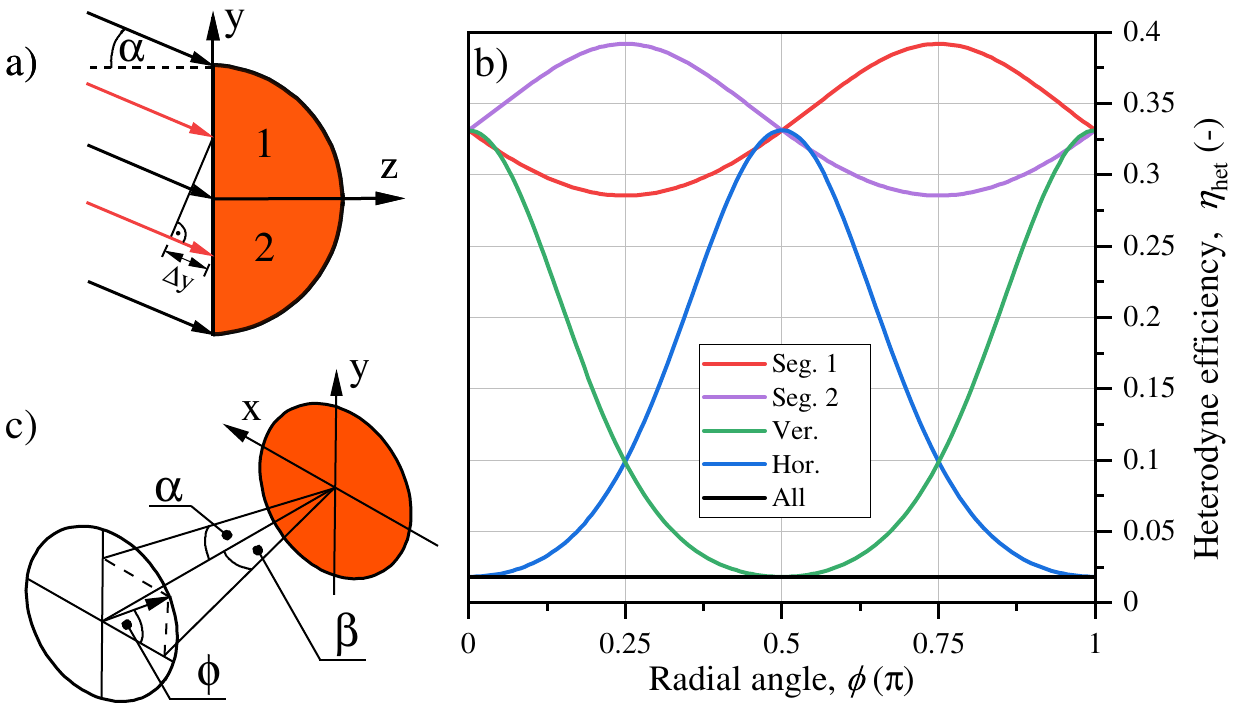}
\par\end{centering}
\centering{}\caption{a) The effect of the signal beam hitting the QPD at a finite angle
$\alpha$ is illustrated. The local oscillator beam is omitted for
clarity. The arrows colored red are used to indicate the resulting
path-length difference $\Delta y$. b) The heterodyne efficiency for
radial angle $\phi$ with $\gamma_{\text{max}}=M_{2}\cdot4.3\,\mu\text{rad}$
is analyzed. Red and purple curves consider a readout of segment 1
and segment 2, respectively. The green curve combines segment 1 and
segment 2, while the blue curve combines segment 1 and 4. The black
curve combines all four segments of the QPD. c) Indication of radial
angle $\phi$ based on a unit circle separated by $1/\tan(\gamma_{\text{max}})$
from the QPD. \label{fig:The-heterodyne-efficiency-radial}}
\end{figure}

Figure \ref{fig:The-heterodyne-efficiency} a) depicts the heterodyne
efficiency under variation of the angle $\alpha_{\text{rx}}$ while
setting $\beta_{\text{rx}}=0$. Hereby, values for beam parameters
and the receiving optics have been taken from \citep{LISAPRFMODEL}.
Note that in case of $\beta_{\text{rx}}=0$, the heterodyne efficiency
of all individual segments is equal so that only the results of segment
1 are plotted. All configurations have their maximum at $\alpha_{\text{rx}}=0$.
However, the heterodyne efficiency decreases much more rapidly with
increased pointing error, i.e. increased $\alpha_{\text{rx}}$, for
all four segments and the vertical segments than for segment 1 and
the horizontal segments. In fact, it can be shown that $\eta_{\text{het},\text{all}}(\beta_{\text{rx}}=0)=\eta_{\text{het},\text{ver}}(\beta_{\text{rx}}=0)\leq\eta_{\text{het},\text{hor}}(\beta_{\text{rx}}=0)=\eta_{\text{het},\text{seg 1}}(\beta_{\text{rx}}=0)$,
see appendix \ref{sec:Appendix-B-Angle}.\\
The degradation in heterodyne efficiency can be explained with respect
to phase differences that arise once light falls on the QPD under
nonzero angles of incidence $\alpha_{\text{rx}}$ and $\beta_{\text{rx}}$,
as evident from Eq. \ref{eq:eta_het_integ}. In this case, additional
phase contributions arise that change along the area of integration,
i.e. along the detector area. The graphical illustration of Fig.
\ref{fig:The-heterodyne-efficiency-radial} a) shows that increasing
angles lead to a dephasing of the coherent beam. The contribution
of $\alpha_{\text{rx}}$ in Eq. \ref{eq:eta_het_integ} leads to a
dephasing along the $y$-axis. Thus, the heterodyne efficiency of
two segments combined along the $x$-axis (horizontal) surpasses the
one of two segments combined along the $y$-axis (vertical), as no
dephasing is present along the $x$-axis. Vice versa, if $\beta_{\text{rx}}$
is varied while $\alpha_{\text{rx}}$ is kept at zero, the heterodyne
efficiency of two segments along the y-axis surpasses the one of two
segments along the x-axis, as evident in Fig. \ref{fig:The-heterodyne-efficiency}
b). Finally, we note that for both scenarios combining all four segments
always results in the lowest heterodyne efficiency, while a single
segment leads to the maximum heterodyne efficiency.

Figure \ref{fig:The-heterodyne-efficiency-radial} b) depicts the
heterodyne efficiency for a variation in $\alpha_{\text{rx}}$ and
$\beta_{\text{rx}}$ expressed in terms of a combined radial angle
$\phi$. Thereby, the angles are related as follows
\begin{align*}
\alpha_{\text{rx}} & =\dfrac{1}{M_{2}}\text{atan}\left(\sin(\phi)\tan(\gamma_{\text{max}})\right),\\
\beta_{\text{rx}} & =\dfrac{1}{M_{2}}\text{atan}\left(\cos(\phi)\tan(\gamma_{\text{max}})\right),
\end{align*}
with a graphical illustration given in Fig. \ref{fig:The-heterodyne-efficiency-radial}
c). Here, we introduced the parameter $\gamma_{\text{max}}$ which
defines the maximum angle of $\alpha_{\text{rx}}$ and $\beta_{\text{rx}}$.
For a maximum angle of $\gamma_{\text{max}}=M_{2}\cdot4.3\,\mu\text{rad}$,
we observe no variation in the heterodyne efficiency once all four
segments are combined due to the rotational symmetry. The heterodyne
efficiencies of two combined segments exhibit a periodicity of $\pi$.
Maxima arise once the angle of incidence is normal to the alignment
of the two QPD segments, in agreement to Fig. \ref{fig:The-heterodyne-efficiency}.
The heterodyne efficiencies of single segments exhibit also a periodicity
of $\pi$. Similar to the two segments, maxima of the single segments
arise once the angle of incidence aligns with the maximum coherent
integration path. For the single segments this path is along the lines
tilted by $\pm\pi/4$ to the $x$ and $y$-axis.\\
For the given $\gamma_{\text{max}}$, the minimum heterodyne efficiency
is observed once all four segments are combined, while the maximum
heterodyne efficiency is present at the single segments. However,
at certain angles $\phi$ the heterodyne efficiency of two combined
segments surpasses the one of a single segment, which prevents a definite
statement on the maximum heterodyne efficiency irrespective of $\phi$.
A definite statement can be obtained when considering the maximum
heterodyne efficiency of the single segments and their combinations.
As shown in appendix \ref{sec:Appendix-B-Angle}, the following inequality
can be found that does not depend on $\gamma_{\text{max}}$ or $\phi$
\begin{equation}
\text{max}(\eta_{\text{het},\text{seg 1}},\eta_{\text{het},\text{seg 2}})\geq\text{max}(\eta_{\text{het},\text{hor}},\eta_{\text{het},\text{ver}})\geq\eta_{\text{het},\text{all}}.\label{eq:ineq_het_eff}
\end{equation}
The inequality in Eq. \ref{eq:ineq_het_eff} is an important result
in the ongoing discussion on the beat note acquisition process. In
the following we will compare these three configurations regarding
the maximum CNR.

\subsection{Noise contributions\label{subsec:Noise-Contributions}}

Dominant noise sources of the long-arm interferometer are shot noise,
relative intensity noise (RIN) and electronic noise \citep{LISAPRFMODEL}.
Stray light coupling noise is usually assumed to be very small and
noise resulting from the pseudo-random noise modulation of the carrier,
which is necessary to enable absolute ranging \citep{Esteban2009},
can be adequately suppressed by suitable selection of the modulation
scheme \citep{sutton2010laser,EuringerTIM2023}. In our investigations
we adapt the model for RIN from \citep{Wissel2022RIN}. To this, we
consider only the 1$f$-RIN contribution since this contribution exceeds
the 2$f$-RIN contributions in the context of LISA by more than two
orders of magnitude \citep{Wissel2023LISA}. Analytic models for shot
noise and electronic noise are found in \citep{Delgado2012}. The
three noise contributions are uncorrelated among each other and result
in a total noise in $\text{V}^{2}/$$\text{Hz}$ of
\begin{align}
N_{0}(f) & =\left(\dfrac{N_{\text{s}}N_{\text{QPD}}\eta_{\text{resp}}}{16}\right)^{2}\left(P_{\text{lo}}^{2}a_{\text{lo,1f}}^{2}+P_{\text{s}}^{2}a_{\text{s,1f}}^{2}\right)\nonumber \\
 & +2e\dfrac{N_{\text{s}}N_{\text{QPD}}\eta_{\text{resp}}}{16}\left(P_{\text{lo}}+P_{\text{s}}\right)+s_{\text{I,seg,el}}^{2}N_{\text{s}}N_{\text{QPD}}\label{eq:Noise_comb}
\end{align}
The first line in Eq. \ref{eq:Noise_comb} represents 1$f$-RIN contribution,
while the first and second term in the second line represent the shot
and electronic noise contribution, respectively. Here, $e$ denotes
the electron charge, $s_{\text{I,seg,el}}$ the electronic noise and
$a_{\text{1f}}$ the 1$f$-RIN coupling \citep{Wissel2023LISA}. Note
that the factor of 16 in denominator of the 1$f$-RIN contribution
and the shot noise arises since the powers $P_{\text{lo}}$ and $P_{\text{s}}$
are defined per QPD. Both, shot noise and RIN, depend on the power
of the local oscillator beam $P_{\text{lo}}$ and the signal beam
$P_{\text{s}}$. Since $P_{\text{s}}/P_{\text{lo}}\lessapprox10^{-6}$
\citep{Wissel2023LISA}, noise contributions arsing from $P_{\text{s}}$
can be safely neglected in Eq. \ref{eq:Noise_comb}. Based on this
approximation an expression for the CNR following Eq. \ref{eq:carrier_power}
and \ref{eq:Noise_comb} can be formed
\begin{align}
\text{C} & \text{NR}\approx\nonumber \\
 & \dfrac{2\eta_{\text{carrier}}^{2}\eta_{\text{resp}}^{2}P_{\text{lo}}P_{\text{s}}(\alpha_{\text{tx}},\beta_{\text{tx}})\eta_{\text{wfe}}\eta_{\text{het}}(M_{2}\alpha_{\text{rx}},M_{2}\beta_{\text{rx}})}{2e\dfrac{16\eta_{\text{resp}}}{N_{\text{s}}N_{\text{QPD}}}P_{\text{lo}}+\left(\eta_{\text{resp}}P_{\text{lo}}a_{\text{lo,1f}}\right)^{2}+\dfrac{16^{2}s_{\text{I,seg,el}}^{2}}{N_{\text{s}}N_{\text{QPD}}}}.\label{eq:CNRapprox}
\end{align}
Based on Eq. \ref{eq:CNRapprox}, only the heterodyne efficiency
and the noise contributions from shot noise and electronic noise depend
on the number of segments considered for the read-out. As shown in
section \ref{subsec:Angle-dependence-of}, given $\alpha_{\text{rx}}=\beta_{\text{rx}}=0$
the heterodyne efficiency is independent of the segments and the number
of segments considered for the read-out. In this case the CNR increases
with increasing the number of segments $N_{\text{s}}$. However, if
light is incident on the QPD at a nonzero angle, the highest heterodyne
efficiency is found for $N_{\text{s}}=1$ , which in turn increases
the CNR. These antagonistic effects are analyzed in the subsequent
section.

\subsection{Angle dependence of CNR\label{subsec:Angle-dependence-CNR}}

\begin{table}
\centering{}\caption{Read-out configurations.\label{tab:Read-out-configurations.}}
\begin{tabular}{ll}
\hline 
\hline Configuration & CNR\tabularnewline
\hline 
One Segment & $\max(\text{CNR}_{\text{seg 1}},\text{CNR}_{\text{seg 2}})$\tabularnewline
Two segment & $\max(\text{CNR}_{\text{hor}},\text{CNR}_{\text{ver}})$\tabularnewline
Four segment & $\text{CNR}_{\text{all}}$\renewcommand{\tabularnewline}{\\ \hline}\tabularnewline
\hline 
\end{tabular}\renewcommand{\tabularnewline}{\\}
\end{table}
\begin{figure*}
\begin{centering}
\includegraphics[width=0.95\linewidth]{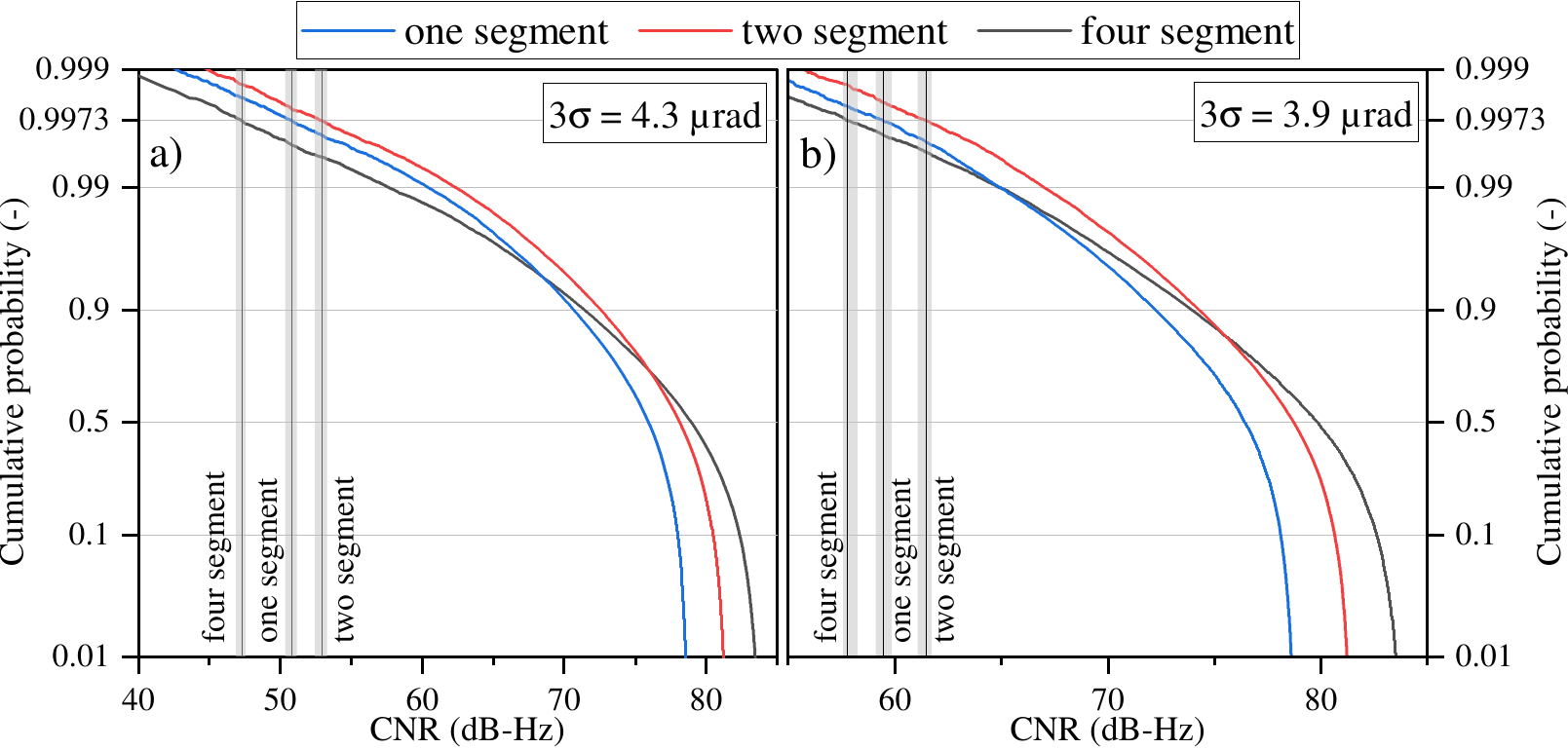}
\par\end{centering}
\centering{}\caption{The CDF of the CNR for a variation of the angles $\alpha_{\text{tx}},\beta_{\text{tx}},\alpha_{\text{rx}}$
and $\beta_{\text{rx}}$ is depicted. Angles are drawn from a normal
distribution with a 3$\sigma$ value of 4.3 $\mu\text{rad}$ and 3.9
$\mu\text{rad}$ in a) and b), respectively. Red and blue curves depict
the CDF for the one and two segment configuration, respectively. The
gray curves represent the CDF for the four segment configuration.
Black vertical lines indicate the 0.997 value of the CDF. Labels at
the bottom of the lines denote the respective configuration. In both
figures the $y$-axis is scaled according to $\ln(y/(1-y))$.\label{fig:Cumulative-probability-of}}
\end{figure*}
The CNR is a crucial parameter during the beat note acquisition process
and details of the implications for the acquisition are discussed
in the subsequent section. From Eq. \ref{eq:CNRapprox} we found
that the CNR depends in particular on the number of segments considered
for the read-out, the received power and the heterodyne efficiency.
The received power itself is a function of the angles $\alpha_{\text{tx}}$
and $\beta_{\text{tx}}$, see Eq. \ref{eq:power_s}, while the heterodyne
efficiency depends on the angles $\alpha_{\text{rx}}$ and $\beta_{\text{rx}}$
and the QPD segments. The number of segments considered for the read-out
is of particular interest. For a perfectly aligned signal beam, i.e.
$\alpha_{\text{rx}}=\beta_{\text{rx}}=0$, the CNR increases with
increasing number of QPD segments. However, for nonzero incidence
angles, i.e. $\alpha_{\text{rx}}\neq0$ and $\beta_{\text{rx}}\neq0$,
the heterodyne efficiency and thus the power of the beat note increases
by reducing the number of segments, cf. Eq. \ref{eq:ineq_het_eff}.
These antagonistic effects shall be analyzed in the following.\\
Since the heterodyne efficiency explicitly depends on the number of
the QPD segments, see Eq. \ref{eq:ineq_het_eff}, we put particular
focus on the three configurations considered in Eq. \ref{eq:ineq_het_eff}.
These three configurations will be denoted as one segment, two segment
and four segment configuration. Thereby, the one segment configuration
considers the maximum CNR of segment 1 and segment 2. The two segment
configuration considers the maximum CNR of the vertical and horizontal
combinations. Finally, the four segment configuration considers the
CNR if the readout of all four segments is combined, see also Tab.
\ref{tab:Read-out-configurations.}. Note that for the one and two
segment configuration only two different read-outs are considered
since the remaining read-outs are identical due to symmetry. The discussion
on the implementation and implications of the maximum operation considered
in the one and two segment configuration has been shifted to section
\ref{subsec:Discussion:-Applicability-of}.\\
The analysis is performed based on a Monte-Carlo simulation of Eq.
\ref{eq:CNRapprox} with 1e5 runs. Thereby, the angles $\alpha_{\text{tx}},\beta_{\text{tx}},\alpha_{\text{rx}}$
and $\beta_{\text{rx}}$ are drawn from a normal distribution with
zero mean and variance $\sigma$. Their uncertainty (3$\sigma$) is
specified in Tab. \ref{tab:Pointing-accuracy.}. Values of the beam
characteristics and the transmitting and receiving optics have been
taken from \citep{LISAPRFMODEL}. Finally, we note that the LISA long-arm
interferometer consists of two QPD pairs operating in hot redundancy
\citep{BarrancoThesis2017}. For the analysis we assume a separate
processing of each QPD pair, while a combination of all QPDs is briefly
discussed in section \ref{subsec:Discussion:-Applicability-of}.\\
Figure \ref{fig:Cumulative-probability-of} depicts the cumulative
probability function (CDF) $F_{\text{CNR}}$ for the three configurations
with respect to the CNR. Black vertical lines indicate the CNR value
of $F_{\text{CNR,lim}}:=0.997$. This value of the CDF means that
99.7\% of all simulations have a CNR that is equal or better than
the value of the associated CNR. For the ongoing discussion this value
will be used as the worst-case CNR. The 1$\sigma$ error of these
vertical lines, which arises from the finite number of simulations,
is indicated by gray boxes. In Fig. \ref{fig:Cumulative-probability-of}
a) the CDF is depicted for an angle uncertainty of 4.3 $\mu\text{rad}$.
Here the two segment configuration exhibits the maximum $F_{\text{CNR,lim}}$
at $\sim52.9$ dB-Hz. This value surpasses $F_{\text{CNR,lim}}$ of
the four segment configuration by roughly 5.6 dB. $F_{\text{CNR,lim}}$
of the one segment configuration is around 2 dB smaller compared to
the two segment configuration. Decreasing the angle uncertainty to
3.9 $\mu\text{rad}$ shifts $F_{\text{CNR,lim}}$ by around 8.6 dB
for the two segment configuration and very similar for the one segment
configuration. For the four segment configuration this shift is even
more prominent with a shift of 10.5 dB, which diminishes the distance
between the two segment and four segment configuration to around 3.6
dB.\\
We can conclude that irrespective of the considered angle uncertainty
(3.9 $\mu\text{rad}$ or 4.3 $\mu\text{rad}$) the two segment configuration
outperforms the one segment configuration and the four segment configuration
based on the worst case CNR represented via $F_{\text{CNR,lim}}$.

\section{Application: Fourier acquisition\label{sec:Application:-Fourier-Acquisition}}

\subsection{Fourier acquisition: basics and model assumptions\label{subsec:Fourier-Acquisition}}

In the previous section we demonstrated that improved CNR levels
of the beat note signal can be obtained through certain QPD read-out
configurations. In the following we want to analyze the implications
of the CNR levels on the LISA beat note acquisition scheme. \\
The beat note acquisition of LISA relies on a Fourier transform acquisition.
The Fourier transform was also used for LISA Pathfinder even during
nominal operation, where amplitude as well as phase information of
a frequency bin were used to stabilize the laser frequency \citep{Hechenblaikner2011LPF,Hechenblaikner2010LPF}.
However, for LISA this is not possible due to the much larger heterodyne
frequency (several MHz instead of kHz for LISA Pathfinder) which changes
dynamically due to orbit evolution and super-imposed frequency offsets
\citep{EESA}. \\
The Fourier transform acquisition considered in the following has
been detailed in \citep{EESA,BrauseThesis2018}. Upon reception,
the beat note signal is digitized with a sampling rate of $f_{\text{s}}=$
80 MHz. Then a fast Fourier transform (FFT) over $L$ samples is
used to convert the signal to the frequency domain, resulting in a
frequency resolution of $f_{\text{res}}=f_{\text{s}}/L$. After taking
the magnitude squared, a peak search algorithm selects the FFT bin
with the maximum signal. It shall be noted that the peak search algorithm
also incorporates the possibility to exclude certain bins from the
search space. This is in particular useful in a laboratory environment
to filter common reference frequencies, however, it will be omitted
in the following.\\
The beat note must be within the receiver bandwidth ($\approx20$
MHz) and below the Nyquist frequency ($f_{s}/2$) to allow a detection
via the FFT acquisition. Since the frequency of the incident beam
is in general not precisely known (uncertainty of around 100 MHz),
the presence of the signal cannot be guaranteed for a Fourier transform
acquisition over a single FFT interval. Consequently, the local
oscillator beam performs a frequency scan over a predefined interval,
in which presence of the beat note can be ensured. For each FFT interval,
the peak search algorithm determines if the maximum of the current
FFT interval supersedes the global maximum of all previous intervals.
If this is the case the maximum of the current interval will be set
as the new global maximum. Once the local oscillator has finished
its scan pattern, the frequency bin associated with the global maximum
is used as initial frequency for the PLL. In the end, the global maximum
then represents the maximum of $M$ number of FFT intervals. If each
FFT interval contains $K$ bins, then the global maximum represents
the maximum over $K_{\text{tot}}=K\cdot M$ frequency bins.

For the ongoing analysis we make the following assumptions \\
(1) Based on a suitable selection of the receiver front-end, we assume
that frequencies above the Nyquist frequency are adequately suppressed
\citep{Barranco2018} and aliasing will be neglected in the following.\\
(2) The finite sample number $L$ per FFT leads to spectral leakage
that may be suppressed via adequate window functions \citep{HeinzelWinFFT2002}.
Once the beat note frequency does not correspond to the center frequency
of a bin, a degradation of the signal bin power due to the spectral
leakage is observed, while the power of adjacent bins is increased.
This well-known behavior of the FFT is in particular addressed via
interpolation algorithms \citep{QuinnFFTinterpol1994,Wang2023FFTCoG}
and thus spectral leakage is omitted in the following. For a worst
case analysis, one may note that the effect is maximized once the
beat note frequency is exactly in between the center frequencies of
two bins. The resulting loss in power of the signal bin is known as
scalloping loss, which is 3.92 dB for a rectangular window \citep{Harris1978}.
This worst case is briefly addressed in the subsequent section.\\
(3) The scan of the local oscillator beam in general ensures that
the beat note is at least once present in one of the $K_{\text{tot}}$
bins. It shall be noted that if presence cannot be ensured, threshold
testing is usually performed to verify its presence. In turn, acquisition
performance strongly depends on the threshold setting. To focus on
the influence of the CNR on the acquisition performance, presence
of the main beat note will be assumed in the following.\\
Moreover, as the scan pattern of the local oscillator beam is not
defined yet and based on assumption (2), we assume that the signal
is present in only one of the $K_{\text{tot}}$ bins. This would either
correspond to a step-scan pattern with non-overlapping FFT intervals
or to a linear scan pattern at a known scan rate. In case of the latter,
effects resulting from the linearly changing frequency can be adequately
compensated by taking into account the discrete chirp-Fourier transform
\citep{XiaChirpDFT2000}. In this way the acquisition can also be
regarded as a single acquisition for $K_{\text{tot}}$ bins. \\
(4) Besides the phase information, the beat note of LISA also enables
the transmission of auxiliary functions, in particular PRN sequences
to enable pseudo-range measurements, data transfer and clock jitter
transfer \citep{sutton2010laser,heinzel2011auxiliary,yamamoto2024experimental}.
Similar concepts are also proposed for the TianQin mission \citep{Xie2023PRNClock,Zeng2023Clock,Xu2024EOM}.
All of these auxiliary functions are realized via low-depth phase
modulations of the carrier, resulting in loss of power for the science
signal, which has been accounted for in Eq. \ref{eq:carrier_power}
by the factor $\eta_{\text{carrier}}$. Around 10\% of the power is
used for the generation of optical sidebands incorporating the clock
jitter. Upon interference with the local oscillator beam, these optical
side-bands form sideband/sideband beat notes one megahertz apart from
the main beat note and well above the noise floor \citep{Esteban:11,EESA,Yamamoto2022}.
Consequently, these sideband/sideband beat notes might lead to false
detection and need to be adequately addressed in the following. \\
PRN sequences and data transmission are realized via a chip modulation,
which consists only of 1\% of the carrier power. Depending on the
chip modulation, the maxima are either in close vicinity to the main
beat note or the sideband/sideband beat notes with noise density maxima
of the order of the shot noise. This and the fact that the chip modulation
can easily be suppressed, e.g., by transmitting a PRN code of constant
zeros, allows us to omit this additional noise contribution in the
following.

\subsection{Probability of detection\label{subsec:Probability-of-detection}}

Based on the assumptions (1) to (4), the influence of the CNR on the
performance of the beat note acquisition scheme can be studied in
an analytical framework delineated in the following. Acquisition performance
is usually evaluated in terms of the PD and the probability of false
alarm. To focus analysis on the CNR, we assumed presence of the main
beat note in one of the bins, see assumption (3), which makes PD and
probability of false alarm $P_{\text{fa}}$ exclusive, i.e. $1=P_{\text{fa}}+P_{\text{D}}$.
Consequently, for the ongoing analysis we focus on the PD, while the
probability of false alarm can be directly retrieved from these results.
\\
For the definition of a correct detection and thus for the evaluation
of the PD, understanding the operating principle of the acquisition
scheme within the phasemeter is crucial: Due to the finite frequency
resolution of the FFT, the output of the beat note acquisition does
not necessarily correspond to the exact beat note frequency. Instead
it represents the initial value for the numerically controlled oscillator
(NCO) of the PLL. A second-order Type 2 PLL, as it is the case for
LISA \citep{EESA}, is then in general capable to achieve the lock.
Hereby, the time to lock depends on the loop parameters, in particular
on the loop bandwidth, and the frequency offset between the beat note
and the initial frequency of the NCO \citep{Gardner2005}. Consequently,
alignment between the frequency resolution of the FFT and the PLL
loop parameters is generally required and will be assumed in the ongoing
analysis. To this end, we do not take into account spectral leakage
as elaborated in section \ref{subsec:Fourier-Acquisition}, assumption
(2). This simplification is especially important as spectral leakage
increases only the noise bins in close vicinity of the beat note.
Consequently, only these bins will be more likely to be selected in
case of spectral leakage. However, based on the phasemeter design,
a selection of such a noise bin does not result in an entirely false
detection but may only increase the final time to lock of the PLL.\\
Based on the assumptions listed in section \ref{subsec:Fourier-Acquisition},
the signal present at the FFT consists of the main beat note, sideband/sideband
beat notes and additive noise. The additive noise is modeled as Gaussian
noise with one-sided noise power density $N_{0}$ as given in Eq.
\ref{eq:Noise_comb}. The power of the sinusoidal beat note is stated
in Eq. \ref{eq:carrier_power}. Expressions on the power of the sideband/sideband
beat notes have been obtained from \citep{Delgado2012}. In this sense,
the input signal represents a real-valued Gaussian random process.
We note that real and imaginary part of each FFT bin again represent
a Gaussian process \citep{Davenport1987}. For an FFT that is normalized
to its sample size, the variance of real and imaginary part is given
by $\sigma_{\text{bin}}^{2}=\tfrac{N_{0}f_{\text{res}}}{4}$, irrespective
if the bin contains the beat note and noise, or only noise. Real and
imaginary parts are then combined in a random variable $S_{\text{bin}}$
using the magnitude squared operation. As shown in \citep{So1999FFTXi2,whalenmcd1995},
the probability density function (PDF) of the normalized variable
$s=S_{\text{bin}}/\sigma_{\text{bin}}^{2}$ for bin $k$ follows 
\begin{equation}
f_{k}(s)=\dfrac{1}{2}e^{-\frac{s+\delta_{k}}{2}}I_{0}(\sqrt{s\delta_{k}}),\label{eq:pdf_chi_sq}
\end{equation}
where $I_{\text{0}}$ denotes the zero order modified Bessel function
and $\delta$ the non-centrality parameter. Noise bins are unbiased
and follow a central $\mathcal{X}_{2}^{2}$-distribution, i.e. $\delta_{k}=0$.
The bins containing the main beat note and the sideband/sideband beat
note are drawn from a non-central $\mathcal{X}_{2}^{2}$-distribution
with non-centrality parameter
\begin{equation}
\delta_{\text{main}}=\tfrac{2\text{CNR}}{f_{\text{res}}}\label{eq:lambda_main}
\end{equation}
and 
\begin{equation}
\delta_{\text{sb}}=\tfrac{J_{1}^{2}(m_{\text{sb}})}{\eta_{\text{carrier}}^{2}}\delta_{\text{main}},\label{eq:lambda_sb}
\end{equation}
respectively. Hereby, the first order Bessel function $J_{1}$ is
a result of the low-depth phase modulation with modulation index $m_{\text{sb}}$
for the side-bands. The modulation index will be assigned with $m_{\text{sb}}=0.45$,
which results in total power of about 10\% for the sideband/sideband
beat notes (5\% per sideband/sideband beat note) \citep{Delgado2012}.

Based on the PDF of signal and noise bins, it is now possible to define
the PD. A correct detection is achieved when the power of the bin
containing the main beat note exceeds all remaining ones. Without
loss of generality, we set bin $j$ as the signal bin, i.e. $\delta_{j}=\delta_{\text{main}}$.
The analytical formulation of the PD is given by the analysis of \citep{Rife1974}
\begin{equation}
P_{\text{D}}=\int_{0}^{\infty}f_{j}(s)\prod_{\underset{k\neq j}{k=0}}^{K_{\text{tot}}-1}F_{k}(s)\,\text{d}s,\label{eq:P_d_intro}
\end{equation}
where $F_{k}(s)$ denotes the CDF of bin $k$. The CDF can be directly
retrieved via integration of Eq. \ref{eq:pdf_chi_sq}. In fact, the
product series contains only two types of CDF, namely the CDF for
the noise bins and the CDF for the sideband/sideband beat notes. We
then find 
\begin{align}
P_{\text{D}} & =\int_{0}^{\infty}\dfrac{1}{2}e^{-\frac{s+\tfrac{2\text{CNR}}{f_{\text{res}}}}{2}}I_{0}(\sqrt{s\tfrac{2\text{CNR}}{f_{\text{res}}}})\cdot\nonumber \\
 & \cdot\left[1-e^{-\frac{s}{2}}\right]^{K_{\text{tot}}-3}\left[1-Q_{1}(\sqrt{\delta_{\text{sb}}},\sqrt{s})\right]^{2}\,\text{d}s,\label{eq:Pd_incl_side}
\end{align}
where the first line is the contribution from the main beat note and
the second line the contribution from all other bins. The function
$Q_{1}$ denotes the Marcum $Q$-function of order one \citep{HelstromMarcQ1992}.
\\
To examine the influence of the sideband/sideband beat notes on the
PD, we will also consider the PD in the case when -- besides the
main beat note -- there are only noise bins, denoted as $P_{\text{D,n}}$.
In this case the PD reads as
\begin{align}
P_{\text{D,n}} & =\int_{0}^{\infty}\dfrac{1}{2}e^{-\frac{s+\tfrac{2\text{CNR}}{f_{\text{res}}}}{2}}I_{0}(\sqrt{s\tfrac{2\text{CNR}}{f_{\text{res}}}})\left[1-e^{-\frac{s}{2}}\right]^{K_{\text{tot}}-1}\,\text{d}s.\label{eq:Pd_no_side_integ}
\end{align}
In fact, following \citep{Steinhardt1985} Eq. \ref{eq:Pd_no_side_integ}
can approximated for high PD values as 
\begin{equation}
P_{\text{D,n}}\approx1-(K_{\text{tot}}-1)\frac{e^{-\frac{\text{CNR}}{2f_{\text{res}}}}}{2}.\label{eq:PD_approx}
\end{equation}
The ongoing analysis is based on the analytical expression of Eq.
\ref{eq:P_d_intro} to Eq. \ref{eq:PD_approx}, however, numerical
simulations have been conducted to justify these results.
\begin{figure}
\begin{centering}
\includegraphics[width=\linewidth]{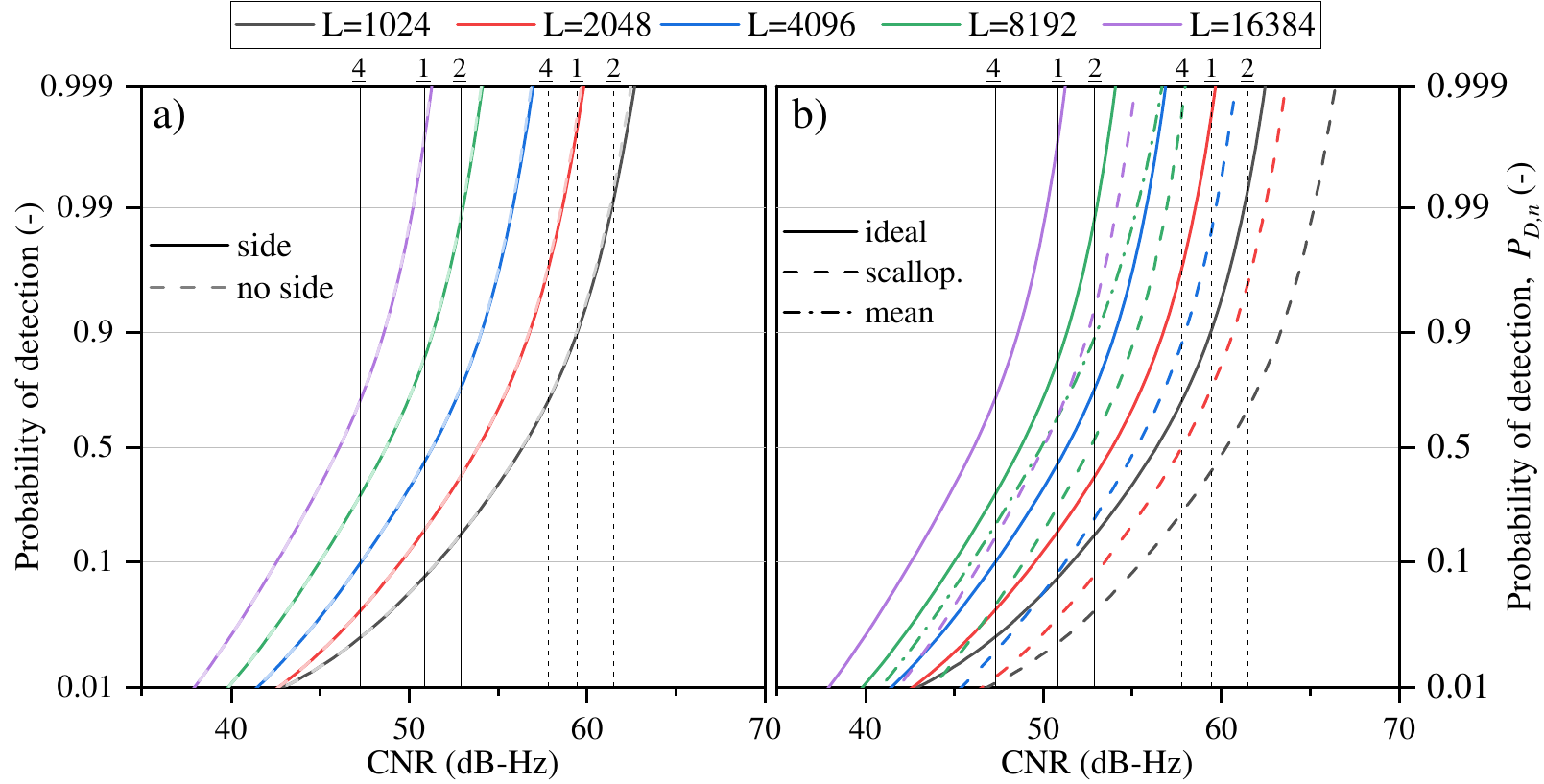}
\par\end{centering}
\centering{}\caption{The PD is plotted over the CNR for different FFT lengths $L$. In
addition, in a) the PD for a signal including (solid lines) and excluding
(dashed lines) sideband/sideband beat notes is depicted. In b) the
effect of scalloping loss is considered (dashed lines) compared to
the case, where the frequency is located at the bin center (solid
lines). Moreover for a FFT length of $L=8192$ also the mean loss
due to spectral leakage is depicted (dashed-dotted line). In both
figures the CNR values for an angle uncertainty of 3.9 $\mu\text{rad}$
(dashed lines) and 4.3 $\mu\text{rad}$ (solid lines) at $F_{\text{CNR,lim}}$
for the one segment ($\text{\ensuremath{\underbar{1}}}$), two segment
($\text{\ensuremath{\underbar{2}}}$) and four segment ($\text{\ensuremath{\underbar{4}}}$)
configuration are indicated by black vertical lines with respective
underlined numbers at the top of each line. In both figures the $y$-axis
is scaled according to $\ln(y/(1-y))$.\label{fig:Probability-of-detection.}}
\end{figure}
\\
First we want to study the influence of the sideband/sideband beat
notes on the PD. Figure \ref{fig:Probability-of-detection.} a) depicts
the PD in presence (Eq. \ref{eq:Pd_incl_side}) and absence (Eq. \ref{eq:Pd_no_side_integ})
of sideband/sideband beat notes by solid and light-dashed lines, respectively.
The lines match near perfectly, revealing that for the given modulation
index the influence of the sideband/sideband beat notes on the PD
is negligible. For the ongoing analysis we will, therefore, restrict
us to the case where sideband/sideband beat notes are absent. Moreover,
this result also shows that the benefit in suppressing the optical
sideband modulation during the beat note acquisition is rather small
from the point of false detection and is mainly due to the increase
of power in the main beat note ($\eta_{\text{carrier}}\rightarrow1$).

\begin{figure}
\begin{centering}
\includegraphics[width=\linewidth]{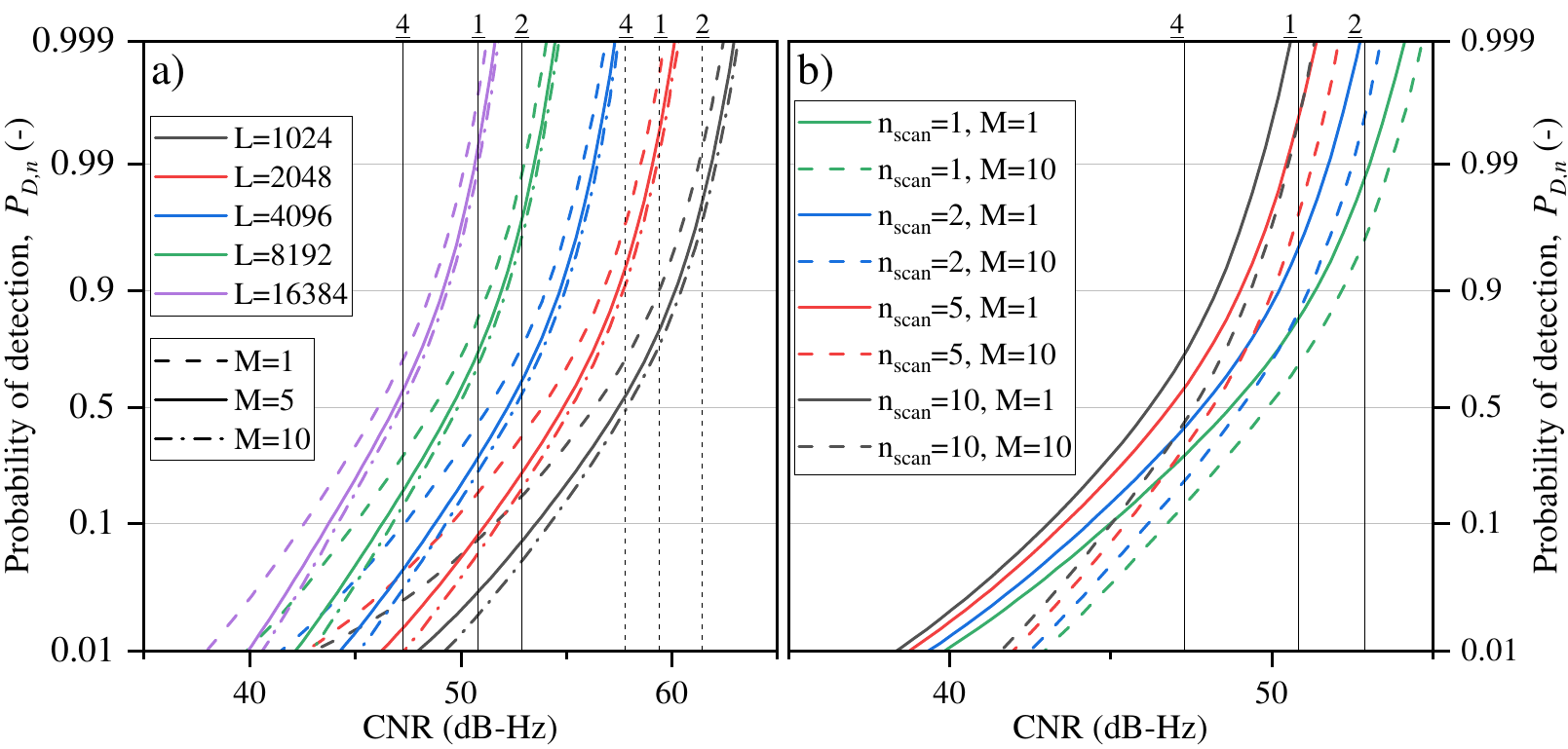}
\par\end{centering}
\centering{}\caption{The influence of multiple FFT intervals on the PD is analyzed in a).
Dashed lines depict the PD for a single interval, while solid and
dashed-dotted lines show the resulting PD for five and ten FFT intervals,
respectively. In b) the influence of multiple scans is analyzed. The
PD for a single scan is depicted by the green plots. The PD for two,
five and ten scans is depicted by the blue, red and gray lines, respectively.
Solid lines represent the PD for a single FFT interval, while dashed
lines show the PD for ten FFT intervals. CNR values for an angle
uncertainty of 3.9 $\mu\text{rad}$ (dashed lines) and 4.3 $\mu\text{rad}$
(solid lines) at $F_{\text{CNR,lim}}$ for the one segment ($\text{\ensuremath{\underbar{1}}}$),
two segment ($\text{\ensuremath{\underbar{2}}}$) and four segment
($\text{\ensuremath{\underbar{4}}}$) configuration are indicated
by black vertical lines with respective underlined numbers at the
top of each line. In both figures the $y$-axis is scaled according
to $\ln(y/(1-y))$. \label{fig:Probability-of-detection.-1}}
\end{figure}

In addition, Fig. \ref{fig:Probability-of-detection.} a) depicts
the PD for different frequency resolutions. Variation of the frequency
resolution $f_{\text{res}}$ is thereby expressed via a variation
of the FFT length $L$. This emphasizes the fact, that in a practical
application, sampling rate and spectrum of interest are fixed. Values
of $L$ are based on previous analysis \citep{MahrdtPhD2014,BrauseThesis2018,EESA}
and the fact that $L$ being an integer power of 2 enables an implementation
via standardized and simple FFT algorithms \citep{BrauseThesis2018}.
Finally, it must be noted that laser frequency noise leads to a broadening
of the laser line width \citep{Elliott1982}, which in turn broadens
the main beat note. Therefore, the frequency resolution and consequently
the FFT length $L$ must be chosen such that the substantial part
of the main beat note power falls inside a single frequency bin of
the FFT. Although for LISA-representative lasers this should generally
still be ensured with an FFT length of $L=16384$, previous studies
focused on smaller FFT lengths. Therefore, we will particularly focus
on $L=8192$, which we will assume as baseline in the following.\\
Importantly, an increase in the FFT length $L$ -- besides the improved
frequency resolution $f_{\text{res}}$ -- also leads to an increasing
number of noise bins $K_{\text{tot}}-1$. In turn, an increasing
number of noise bins decreases the PD, as apparent from Eq. \ref{eq:PD_approx}.
Consequently, although the contribution of the frequency resolution
$f_{\text{res}}$ and the CNR on the PD are theoretically equal, as
apparent from Eq. \ref{eq:Pd_no_side_integ}, in a practical application
this is not the case and the PD is most sensitive to the CNR. This
can also be observed in Fig. \ref{fig:Probability-of-detection.}
a), which depicts the PD for a single FFT interval covering a spectrum
of 20 MHz. Frequency resolution of the purple curve is increased by
a factor of 16 compared to the gray curve. This would correspond to
a shift of 12 dB between both curves. However, the actual shift at
high values is only around 11.4 dB due to the influence of the increased
number of noise bins.\\
In Fig. \ref{fig:Probability-of-detection.} b) the effect of scalloping
loss for a rectangular window is illustrated. Irrespective of the
FFT length, it leads to a shift of the curves by value of the scalloping
loss (3.92 dB), which is a direct consequence from Eq. \ref{eq:Pd_no_side_integ}.
In addition, for a FFT length of $L=8192$, also the PD considering
the mean power loss due to spectral leakage has been taken into account.
Based on the operating principle of the acquisition scheme within
the phasemeter, spectral leakage has been incorporated in Eq. \ref{eq:Pd_no_side_integ}
via a decrease in CNR in the signal bin. The resulting PD is shifted
by around 1 dB from the ideal PD line, i.e. when the frequency is
located at the bin center, for PD values of $\lessapprox0.5$. Towards
higher PD values the influence of the spectral leakage increases to
around 2.6 dB at PD $=0.999$.\\
PD curves are brought into perspective with the different read-out
schemes discussed in section \ref{subsec:Angle-dependence-CNR} via
the black vertical lines indicating the value of $F_{\text{CNR,lim}}$
for an angle uncertainty of $3.9\,\mu\text{rad}$ (dashed lines)
and $4.3\,\mu\text{rad}$ (solid lines). In case of an angle uncertainty
of $4.3\,\mu\text{rad}$, only the one and two segment configuration
with an FFT length of $L=16384$ achieve a high PD ($>99$\%). For
$3.9\,\mu\text{rad}$ the two segment configuration guarantees a high
PD over all FFT lengths considered in the analysis if the beat note
frequency is located at the bin center, while the four segment configuration
requires an FFT length of $L\geq4096$. Once the signal is located
in between two bins, a high PD cannot be guaranteed for an angle uncertainty
of $4.3\,\mu\text{rad}$ irrespective of the configuration, while
for an angle uncertainty of $3.9\,\mu\text{rad}$ a high PD is obtained
at an FFT length of $L\geq4096$ and $L\geq8192$ for the two and
four segment configuration, respectively. \\
The majority of these results retains its validity once multiple FFT
intervals are considered. As depicted in Fig. \ref{fig:Probability-of-detection.-1}
a), at high PD values the influence of an increasing number of FFT
intervals $M$ is rather small, yielding a CNR shift of around $0.5$
dB for $M=5$ at $P_{\text{D,n}}=0.99$. An increasing influence is
observed towards lower PD values. As a baseline case we may consider
a receiver bandwidth of 20 MHz and an uncertainty of laser frequency
of 100 MHz, necessitating a laser frequency scan over approximately
5 FFT intervals. In this case for an angle uncertainty of $4.3\,\mu\text{rad}$
and an FFT length of $L=8192$, the PD of the two segment configuration
is around 0.97, while the PD of the four segment configuration is
only at around 0.17. This result clearly highlights the performance
improvement obtained through the two segment configuration. However,
even the PD of the two segment configuration may not be suitable and
a greater FFT length might be required. 
\\
Another method to increase the PD is to perform multiple scans over
the spectrum of interest. It shall be noted that variations of the
main beat note due to the relative SC motion does not affect this
method. The relative velocity of the SC is around 15 m/s \citep{OttoThesis2015}.
This leads to slowly varying Doppler shifts negligible on the time
scale of the FFT acquisition. Consequently, the signal will remain
in the same bin for each scan. Combining the scans, there will be
$n_{\text{scan}}K_{\text{tot}}$ bins and the main beat note will
be contained in $n_{\text{scan}}$ bins, where $n_{\text{scan}}$
equals the number of scans. Again selecting the main beat note based
on the maximum bin power, the PD in absence of the sideband/sideband
beat notes can still be calculated via Eq. \ref{eq:Pd_no_side_integ}
with the following adaptions: The PDF of the main beat note needs
to be replaced with the maximum PDF of $n_{\text{scan}}$ beat notes,
which is given by $n_{\text{scan}}f_{j}(s)(1-Q_{1}(\sqrt{\delta_{\text{main}}},\sqrt{s}))^{n_{\text{scan}}-1}$.
In the same way also the number of noise bins increases to $n_{\text{scan}}(K_{\text{tot}}-1)$.
Figure \ref{fig:Probability-of-detection.-1} b) depicts the resulting
PD for a fixed FFT length of $L=8192$. An increase in $n_{\text{scan}}$
results in an increased PD. The performance increase from one to five
signal bins is thereby comparable with switching from the read-out
scheme of the one segment configuration to the two segment configuration
for an angle uncertainty of 4.3 $\mu\text{rad}$. Importantly, the
influence of multiple scans is most relevant at high PD values. Already
two scans can ensure $P_{\text{D,n}}>0.99$ for the two segment configuration
for an FFT interval of $M=10$, an FFT length of $L=8192$, and an
angle uncertainty of 4.3 $\mu\text{rad}$. However, multiple scans
result in an increased acquisition time.\\
In summary, analysis of the PD revealed that the CNR values, where
the CDF reaches 99.7\% for an angle uncertainty of 3.9 $\mu\text{rad}$
and 4.3 $\mu\text{rad}$, are in a critical region, where depending
on the FFT acquisition parameters, high PD values ($>0.99$) cannot
be ensured. Increasing the FFT length $L$ or the number of scans
over the spectrum of interest can increase the PD. This, however,
comes at the price of higher processing load and an increased processing
time. Moreover, it has been shown that the PD is most sensitive to
the CNR compared to aforementioned FFT acquisition parameters. In
consequence, in particular for an angle uncertainty of 4.3 $\mu\text{rad}$,
the increase in CNR provided by the two segment configuration led
to a strong PD improvement compared to the four segment configuration.
Irrespective of the angle uncertainty, these results highlight the
advantages and necessity of a well-considered decision on the read-out
scheme.\\
In addition, the analysis revealed the importance of reducing the
errors in the angle uncertainty. The shift in the uncertainty budget
for 3.9 $\mu\text{rad}$ instead of 4.3 $\mu\text{rad}$, already
shifts the black vertical lines to the right by about 9 dB. This
statement highlights the large payoff to be gained by slightly reducing
the various errors that enter the uncertainty budget.

\subsection{Discussion\label{subsec:Discussion:-Applicability-of}}

For the considered angle uncertainties of 3.9 $\mu\text{rad}$ and
4.3 $\mu\text{rad}$, the two segment configuration outperforms the
one and four segment configurations in terms of CNR. The previous
section highlighted the importance of the CNR, and consequently also
the adequate selection of the read-out configurations, on the FFT
acquisition. In the following, the implementation of the read-out
configurations in terms of the FFT acquisition shall be addressed.
\\
One and two segment configuration rely on the maximum operation as
defined in section \ref{subsec:Angle-dependence-CNR}. In terms
of the FFT acquisition, the implementation of the maximum operation
of two different read-outs necessitates two separate FFT acquisition
schemes, where at the end the maximum of both acquisition schemes
is considered. Taking into account the assumptions made in section
\ref{subsec:Fourier-Acquisition}, the expectation of the output then
equals the maximum CNR of both read-outs. \\
Since the maximum operation and thus a second FFT is only considered
for the one and two segment configuration, naturally the question
arises about a second FFT (1) or and increased frequency resolution
(2) for the four segment configuration. This shall be briefly discussed
in the following: 

(1) A second FFT could be used to consider the read-out of additional
QPDs operating in hot redundancy. Hereby, the output of both could
be used to form a maximum operation as it is done for the one and
two segment configuration. Since the CNR is equal for both read-outs
the PD is described by Fig. \ref{fig:Probability-of-detection.-1}
b) for $n_{\text{scan}}=2$. The curve for $n_{\text{scan}}=2$ is
shifted by around 1 to 2 dB compared to $n_{\text{scan}}=1$. This
is well below the CNR increase of the two segment configuration compared
to the four segment (3.7 dB for an angle uncertainty of 3.9 $\mu\text{rad}$
and 5.6 dB for 4.3 $\mu\text{rad}$). \\
In fact, the second output could also be used to form a summation
of both FFTs after taking the magnitude squared. The PD is then obtained
via Eq. \ref{eq:P_d_intro} by noting that the bin power follows a
$\mathcal{X}_{4}^{2}$-distribution with non-centrality parameter
$2\delta$. This type of summation corresponds to an incoherent summation
and has been detailed \citep{MahrdtPhD2014}.

(2) It is well-known that the operations of a FFT scale with $\mathcal{O}(L\log L)$
\citep{Cooley1965}. Consequently for large $L$, instead of a second
FFT, a FFT with roughly double the frequency resolution could be used.
Based on the findings in section \ref{subsec:Probability-of-detection}
this would shift the PD curve by <3dB. \\
Consequently, irrespective of the angle uncertainty (3.9 $\mu\text{rad}$
or 4.3 $\mu\text{rad}$) the two segment configuration would still
outperform the four segment configuration.

\section{Conclusion}

In this paper we analyzed the coherent acquisition architecture for
LISA, determining the feasibility range and optimizing the driving
design parameters. Two major aspects are highlighted: the CNR of the
beat note signal read from the QPD for different read-out configurations,
and the probability of detection for the baseline acquisition scheme
with focus on the CNR.\\
The CNR is governed by the power of the received beam, the heterodyne
efficiency of the interference signal, and the noise affecting the
signal and its phase readout. These quantities depend on the angles
of the transmitted beam and of the received beam with respect to the
line-of-sight in between two SC, where the residual pointing errors
are incurred from the preceding spatial acquisition phase. The heterodyne
efficiency plays a special role as it is the only quantity that depends
on the coherence between local and received beam. Detailed analysis
on the heterodyne efficiency revealed that single segments of a QPD
are less sensitive to a pointing error compared to two combined segments.
In particular, it could be shown that the maximum heterodyne efficiency
of two adjacent single segments (one segment configuration) always
outperforms the maximum CNR of any combination of two adjacent segments
(two segment configuration). The heterodyne efficiency once all four
segments of a QPD are combined (four segment configuration) is in
any case most sensitive to pointing errors.\\
Considering this effect, and also the dependency of the received beam
power on the angle of the transmitted beam, we find from Monte Carlo
simulations that the expected minimum CNR of the beat note signal
is highest for the two segment configuration. In this configuration
and for 4.3 $\mu\text{rad}$ pointing error (3$\sigma)$, the expected
minimum CNR exceeds the four segment configuration by 5.6 dB. The
minimum CNR for all configurations was defined from the respective
cumulative probability at the 99.73\% level, meaning that 99.73\%
of simulations yield better or equal CNR.\\
Application and influence of the configurations on the beat note acquisition
has been analyzed in the second part of the paper. The minimum CNR
is the most essential parameter for calculating the probability of
detecting the beat note for which an FFT peak search algorithm is
baselined for LISA. The PD is also strongly impacted by the number
of points used in the FFT, i.e., the measurement time, as well as
the spectral leakage, which is analyzed in terms of the scalloping
loss, while false detection due to the optical sidebands are found
to have only negligible effect. Furthermore, we considered the impact
of the number of frequency search intervals on the PD, noting that
the uncertainty of laser frequency ($\approx100$ MHz) is much higher
than the phasemeter bandwidth ($\approx20$ MHz). Taking all these
effects into account and assuming the minimum CNR found for each configuration,
we find that for the two segment configuration the PD is around 97\%
while for the four segment configuration it is only around 17\%. Here,
we assumed an angular uncertainty of 4.3 $\mu\text{rad}$ and 8192
FFT points, while scalloping loss has been neglected. We also showed
the possibility of increasing the PD by performing multiple scans
across the same frequency range, but this comes at the expense of
longer measurement times.\\
Finally, analyses have been extended taking into account that the
implementation of the one and two segment configurations necessitates
a second FFT acquisition scheme. It was shown that when considering
similar additions for the four segment configuration, e.g. by doubling
the FFT length, the two segment configuration still outperforms the
four segment configuration in terms of minimum CNR.\\
The analysis approach and models presented in this paper may be used
as the framework for defining the architecture and design parameters
of the coherent acquisition phase for LISA or other interferometry
missions in space. 
\begin{acknowledgments}
The authors thank T. Ziegler, O. Mandel, S. Delchambre, P. Gath and
P. Voigt for their support and fruitful discussions. \\
This work was supported by funding from the Max-Planck-Institut für
Gravitationsphysik (Albert-Einstein-Institut), based on a grant by
the Deutsches Zentrum für Luft- und Raumfahrt (DLR). The work was
supported by the Bundesministerium für Wirtschaft und Klimaschutz
based on a resolution of the German Bundestag (Project Ref. Number
50 OQ 1801).
\end{acknowledgments}

\appendix

\section{Heterodyne efficiency\label{sec:Appendix-A-Heterodyne}}

In the following, we derive an expression for the heterodyne efficiency
$\eta_{\text{het}}$ for a single segment, two combined segments and
four combined segments of a QPD. Following the definition in Eq. \ref{eq:intro_het_eff},
the heterodyne efficiency in the ORF is given by \begin{widetext} \small
\begin{align}
\eta_{\text{het}} & =\dfrac{\left|\iint E_{\text{s}}^{\ast}(\alpha_{\text{tx}},\beta_{\text{tx}},\alpha_{\text{rx}},\beta_{\text{rx}})E_{\text{lo}}\,\text{d}S\right|^{2}}{\iint\left|E_{\text{s}}\right|^{2}\,\text{d}S\,\iint\left|E_{\text{lo}}\right|^{2}\,\text{d}S}\nonumber \\
 & =\dfrac{(\tfrac{1}{2}c\epsilon_{0})^{2}|E_{\text{s}}^{0}(\alpha_{\text{tx}},\beta_{\text{tx}})|^{2}|E_{\text{lo}}^{0}|^{2}}{(N_{\text{s}}/4)^{2}P_{\text{s}}P_{\text{lo}}}\left|\iint\exp(\text{i}k_{\text{s}}(x\sin(M_{2}\beta_{\text{rx}})-y\sin(M_{2}\alpha_{\text{rx}})\cos(M_{2}\beta_{\text{rx}})))\exp\left(-\dfrac{x^{2}+y^{2}}{W_{0}^{2}}\right)\,\text{d}\tilde{S}\right|^{2}\nonumber \\
 & =\dfrac{c\epsilon_{0}|E_{\text{lo}}^{0}|^{2}}{2\pi(N_{\text{s}}/4)^{2}R_{\text{min}}^{2}P_{\text{lo}}}\left|\iint\exp(-(x^{2}/W_{0}^{2}-\text{i}k_{\text{s}}x\sin(M_{2}\beta_{\text{rx}})))\exp(-(y^{2}/W_{0}^{2}+\text{i}k_{\text{s}}y\sin(M_{2}\alpha_{\text{rx}})\cos(M_{2}\beta_{\text{rx}})))\,\text{d}\tilde{S}\right|^{2}.\label{eq:calc_eta_het}
\end{align}
\normalsize Hereby, $\text{d}\tilde{S}$ indicates that, in contrast
to $\text{d}S$, integration is only performed over the area that
is not clipped by the aperture with radius $R_{\text{min}}$ (see
section \ref{subsec:Generic-Model}). We note that the power of the
local oscillator beam (Eq. \ref{eq:def_local_beam}) at the QPD is
given by: \small
\begin{align*}
P_{\text{lo}} & =\tfrac{1}{2}c\epsilon_{0}\underset{\text{all}}{\iint}\left|E_{\text{lo}}\right|^{2}\,\text{d}S=\dfrac{1}{2}c\epsilon_{0}\int_{0}^{R_{\text{QPD}}}\int_{0}^{2\pi}\left|E_{\text{lo}}^{0}\right|^{2}\cdot e^{-2(r/W_{0})^{2}}r\,\text{d}\varphi\text{d}r=\pi c\epsilon_{0}\left|E_{\text{lo}}^{0}\right|^{2}\dfrac{W_{0}^{2}}{4}\left[1-e^{-2(R_{\text{QPD}}/W_{0})^{2}}\right].
\end{align*}
\normalsize Consequently, the amplitude reads as \small
\begin{align}
E_{\text{lo}}^{0} & =\dfrac{1}{W_{0}\sqrt{2\pi c\epsilon_{0}}}\sqrt{P_{\text{lo}}}N(R_{\text{QPD}}/W_{0}),\:\:\text{with}\:N(x):=2^{3/2}\left(1-e^{-2x^{2}}\right)^{-1/2}.\label{eq:def_E_lo}
\end{align}
\normalsize Inserting Eq. \ref{eq:def_E_lo} in Eq. \ref{eq:calc_eta_het}
yields \small
\begin{equation}
\eta_{\text{het}}=\left(\dfrac{2N(R_{\text{QPD}}/W_{0})}{N_{\text{s}}W_{0}\pi R_{\text{min}}}\right)^{2}\left|\iint\exp(-(x^{2}/W_{0}^{2}-\text{i}k_{\text{s}}x\sin(M_{2}\beta_{\text{rx}})))\exp(-(y^{2}/W_{0}^{2}+\text{i}k_{\text{s}}y\sin(M_{2}\alpha_{\text{rx}})\cos(M_{2}\beta_{\text{rx}})))\,\text{d}\tilde{S}\right|^{2}.\label{eq:eta_het_intro}
\end{equation}
\normalsize Finally, expanding the term in the first exponential
function according to \small
\begin{align*}
\eta_{\text{het}} & =\left(\dfrac{2N(R_{\text{QPD}}/W_{0})}{N_{\text{s}}W_{0}\pi R_{\text{min}}}\right)^{2}\exp\left(-\frac{1}{2}W_{0}^{2}k_{s}^{2}\sin^{2}\left(M_{2}\beta_{\text{rx}}\right)\right)\cdot\\
 & \cdot\left|\iint\exp\left(-\left(x/W_{0}-\frac{\text{i}}{2}W_{0}k_{s}\sin\left(M_{2}\beta_{\text{rx}}\right)\right)^{2}\right)\exp\left(-y^{2}/W_{0}^{2}-\text{i}k_{\text{s}}y\sin\left(M_{2}\alpha_{\text{rx}}\right)\cos\left(M_{2}\beta_{\text{rx}}\right)\right)\,\text{d}\tilde{S}\right|^{2}
\end{align*}
\normalsize enables the integration in Cartesian coordinates i.e.
$\text{d}\tilde{S}\rightarrow\text{d}x\text{d}y$. Therefore, we introduce
the following notation
\begin{equation}
\text{erf}(a,b):=\text{erf}(a)-\text{erf}(b),\:\text{with}\:\text{erf}(x)={\displaystyle \frac{2}{\sqrt{\pi}}\int_{0}^{x}e^{-t^{2}}\,\mathrm{d}t,}\label{eq:def_erf_com}
\end{equation}
which results in \small
\begin{align}
\eta_{\text{het}} & =\left(\dfrac{N(R_{\text{QPD}}/W_{0})}{N_{\text{s}}\sqrt{\pi}R_{\text{min}}}\right)^{2}\exp\left(-\frac{1}{2}W_{0}^{2}k_{s}^{2}\sin^{2}\left(M_{2}\beta_{\text{rx}}\right)\right)\cdot\nonumber \\
 & \cdot\left|\int_{y^{-}}^{y^{+}}\text{erf}\left(\frac{x^{+}}{W_{0}}-\frac{1}{2}iW_{0}k_{s}\sin\left(M_{2}\beta_{\text{rx}}\right),\frac{x^{-}}{W_{0}}-\frac{1}{2}iW_{0}k_{s}\sin\left(M_{2}\beta_{\text{rx}}\right)\right)\exp\left(-y^{2}/W_{0}^{2}-\text{i}k_{\text{s}}y\sin\left(M_{2}\alpha_{\text{rx}}\right)\cos\left(M_{2}\beta_{\text{rx}}\right)\right)\,\text{d}y\right|^{2}.\label{eq:eta_het_eval}
\end{align}
\normalsize Hereby, selection of $y^{\pm}$ and $x^{\pm}$ according
to Tab. \ref{tab:QPD-configurations.} enables to calculate the heterodyne
efficiency $\eta_{\text{het}}$ for a single segment, two combined
segments and all four combined segments of a QPD. \end{widetext}

\section{Angle dependence of heterodyne efficiency\label{sec:Appendix-B-Angle}}

In this appendix we evaluate the heterodyne efficiency for various
segments when the signal beam hits the QPD at a certain angle of incidence.
To this end, we will adopt the notation specified in Tab. \ref{tab:QPD-configurations.}.
The evaluation will be performed based on Eq. \ref{eq:eta_het_eval}.
However, to reduce the amount of lengthy expressions, we will focus
only on the terms which differ among the segments and segment configurations,
and denote this parameter as $\eta_{\text{het}}^{\prime}$. Clearly,
the results also hold for $\eta_{\text{het}}$. Based on Eq. \ref{eq:eta_het_eval},
we will define $\eta_{\text{het}}^{\prime}$ as \begin{widetext} \small
\begin{align}
\eta_{\text{het}}^{\prime} & :=\dfrac{1}{N_{\text{s}}^{2}}\left|\int_{y^{-}}^{y^{+}}\text{erf}\left(\frac{x^{+}}{W_{0}}-\frac{1}{2}iW_{0}k_{s}\sin\left(M_{2}\beta_{\text{rx}}\right),\frac{x^{-}}{W_{0}}-\frac{1}{2}iW_{0}k_{s}\sin\left(M_{2}\beta_{\text{rx}}\right)\right)\exp\left(-y^{2}/W_{0}^{2}-\text{i}k_{\text{s}}y\sin\left(M_{2}\alpha_{\text{rx}}\right)\cos\left(M_{2}\beta_{\text{rx}}\right)\right)\,\text{d}y\right|^{2}.\label{eq:eta_prime}
\end{align}
\end{widetext} \normalsize Moreover, the following identities of
definition \ref{eq:def_erf_com} will be particularly useful
\begin{align}
\text{erf}\left(a+bi,-a+bi\right) & \in\mathbb{R}\label{eq:id_error_func_real}\\
\mathfrak{R}\{\text{erf}\left(a+bi,bi\right)\} & =\dfrac{1}{2}\,\text{erf}\left(a+bi,-a+bi\right)\label{eq:id_error_func_2}\\
\left(\text{erf}\left(a+bi,bi\right)\right)^{\ast} & =\text{erf}(bi,-a+bi)\label{eq:id_error_func_3}\\
\text{erf}\left(a,-a\right) & =2\,\text{erf}(a)\label{eq:id_error_func_4}
\end{align}
with $a,b\in\mathbb{R}$. Proofs for the identities are straightforward
by expressing the complex error function in definition \ref{eq:def_erf_com}
through line segments as done in \citep{Zaghloul2011}.\\
First we will consider a variation of the angle of incidence only
along one axis and set $\beta_{\text{rx}}=0$. Starting with two combined
segments ($N_{\text{s}}=2$) along the horizontal axis we have $x^{+}=-x^{-}$.
Therefore, taking into account Eq. \ref{eq:id_error_func_4}, $\eta_{\text{het}}^{\prime}$
as defined in Eq. \ref{eq:eta_prime} yields \begin{widetext} \small
\begin{equation}
\eta_{\text{het},\text{hor}}^{\prime}=\left|\int_{0}^{R_{\text{min}}}\text{erf}\left(\frac{\sqrt{R_{\text{min}}^{2}-y^{2}}}{W_{0}}\right)\exp\left(-y^{2}/W_{0}^{2}\right)\exp\left(-\text{i}k_{\text{s}}y\sin(M_{2}\alpha_{\text{rx}})\cos(M_{2}\beta_{\text{rx}})\right)\,\text{d}y\right|^{2}.\label{eq:eta_het_1_4_no_beta_new}
\end{equation}
On the other hand for the vertical segments (segments 1+2) in case
of $\beta_{\text{rx}}=0$ we have
\begin{align}
\eta_{\text{het},\text{ver}}^{\prime} & =\dfrac{1}{4}\left|\int_{-R_{\text{min}}}^{R_{\text{min}}}\text{erf}\left(\frac{\sqrt{R_{\text{min}}^{2}-y^{2}}}{W_{0}}\right)\exp\left(-y^{2}/W_{0}^{2}\right)\exp\left(-\text{i}k_{\text{s}}y\sin(M_{2}\alpha_{\text{rx}})\cos(M_{2}\beta_{\text{rx}})\right)\,\text{d}y\right|^{2}\label{eq:eta_het_2_1_no_beta_new_pre}\\
 & =\left|\int_{0}^{R_{\text{min}}}\text{erf}\left(\frac{\sqrt{R_{\text{min}}^{2}-y^{2}}}{W_{0}}\right)\exp\left(-y^{2}/W_{0}^{2}\right)\cos\left(k_{\text{s}}y\sin(M_{2}\alpha_{\text{rx}})\cos(M_{2}\beta_{\text{rx}})\right)\,\text{d}y\right|^{2}.\label{eq:eta_het_2_1_no_beta_new}
\end{align}
\end{widetext} \normalsize In the last line we applied Euler's formula
and noted that the only odd function in \ref{eq:eta_het_2_1_no_beta_new_pre}
is the sine. From Eq. \ref{eq:eta_het_1_4_no_beta_new} and Eq. \ref{eq:eta_het_2_1_no_beta_new}
it is now clear that: $\eta_{\text{het},\text{ver}}^{\prime}(\beta_{\text{rx}}=0)\leq\eta_{\text{het},\text{hor}}^{\prime}(\beta_{\text{rx}}=0)$
and equivalently $\eta_{\text{het},\text{ver}}(\beta_{\text{rx}}=0)\leq\eta_{\text{het},\text{hor}}(\beta_{\text{rx}}=0)$.
Moreover, in a similar way it can be shown that $\eta_{\text{het},\text{ver}}(\alpha_{\text{rx}}=0)\geq\eta_{\text{het},\text{hor}}(\alpha_{\text{rx}}=0)$.
We also note that in case of $\beta_{\text{rx}}=0$ a single segment
results in the identical heterodyne efficiency as $\eta_{\text{het},\text{hor}}$,
while combining all four segments leads to the heterodyne efficiency
of $\eta_{\text{het},\text{ver}}$. Consequently, we have $\eta_{\text{het},\text{seg }j}(\alpha_{\text{rx}}=0)\geq\eta_{\text{het},\text{all}}(\alpha_{\text{rx}}=0)$
and $\eta_{\text{het},\text{seg }j}(\beta_{\text{rx}}=0)\geq\eta_{\text{het},\text{all}}(\beta_{\text{rx}}=0)$
with $j\in\{\text{1,2}\}$.

In the following, we will examine the heterodyne efficiency under
variation of $\alpha_{\text{rx}}$ and $\beta_{\text{rx}}$. At first
we analyze $\eta_{\text{het}}^{\prime}$ for two combined segments
($N_{\text{s}}=2$) compared to four combined segments ($N_{\text{s}}=4$).
For the two segments we will consider the horizontal segments, however,
the procedure can easily be transferred to the vertical segments.
Based on Eq. \ref{eq:eta_prime}, the combined heterodyne efficiency
for the horizontal segments is given by \small
\begin{align}
\eta_{\text{het},\text{hor}}^{\prime} & =\left|\int_{0}^{R_{\text{min}}}A_{\text{re}}\exp\left(-\text{i}k_{\text{s}}y\sin(M_{2}\alpha_{\text{rx}})\cos(M_{2}\beta_{\text{rx}})\right)\,\text{d}y\right|^{2},\label{eq:eta_hor_a_b}
\end{align}
with
\begin{align}
A_{\text{re}} & :=\dfrac{1}{2}e^{-y^{2}/W_{0}^{2}}\text{erf}(\sqrt{R_{\text{min}}^{2}-y^{2}}/W_{0}-\frac{1}{2}iW_{0}k_{s}\sin(M_{2}\beta_{\text{rx}}))\nonumber \\
 & -\dfrac{1}{2}e^{-y^{2}/W_{0}^{2}}\text{erf}(-\sqrt{R_{\text{min}}^{2}-y^{2}}/W_{0}-\frac{1}{2}iW_{0}k_{s}\sin(M_{2}\beta_{\text{rx}})).\label{eq:def_A_re}
\end{align}
\normalsize According to Eq. \ref{eq:id_error_func_real}, $A_{\text{re}}$
is real. Performing similar steps as in Eq. \ref{eq:eta_het_2_1_no_beta_new},
the heterodyne efficiency once all four segments are combined is given
by \small
\begin{align}
\eta_{\text{het},\text{all}}^{\prime} & =\left|\int_{0}^{R_{\text{min}}}A_{\text{re}}\cos\left(k_{\text{s}}y\sin(M_{2}\alpha_{\text{rx}})\cos(M_{2}\beta_{\text{rx}})\right)\,\text{d}y\right|^{2}.\label{eq:eta_all}
\end{align}
\normalsize Comparison of Eq. \ref{eq:eta_hor_a_b} and Eq. \ref{eq:eta_all}
then yields $\eta_{\text{het},\text{hor}}\geq\eta_{\text{het},\text{all}}$.
Due to rotational symmetry of the problem the statement also holds
if instead of the two horizontal segments the two vertical segments
are combined. We can summarize that the heterodyne efficiency of two
combined adjacent segments outperforms the heterodyne efficiency once
all four segments are combined.

Finally, we want to compare the heterodyne efficiency of a single
segment to the one of two combined adjacent segments. Therefore, we
will introduce the following substitutions \small
\begin{align}
A_{\text{re}}+\text{i}A_{\text{im}} & =-\dfrac{1}{2}e^{-y^{2}/W_{0}^{2}}\text{erf}(-\frac{1}{2}iW_{0}k_{s}\sin(M_{2}\beta_{\text{rx}}))\nonumber \\
+\dfrac{1}{2}e^{-y^{2}/W_{0}^{2}} & \text{erf}(\sqrt{R_{\text{min}}^{2}-y^{2}}/W_{0}-\frac{1}{2}iW_{0}k_{s}\sin(M_{2}\beta_{\text{rx}})),\label{eq:def_Are}
\end{align}
\normalsize and \small
\begin{equation}
B_{\text{re}}+\text{i}B_{\text{im}}=\exp\left(-\text{i}k_{\text{s}}y\sin(M_{2}\alpha_{\text{rx}})\cos(M_{2}\beta_{\text{rx}})\right),\label{eq:def_Bre}
\end{equation}
\normalsize with $A_{\text{re}},A_{\text{im}},B_{\text{re}},B_{\text{im}}\in\mathbb{R}$.
Note that based on Eq. \ref{eq:id_error_func_2}, the definition of
$A_{\text{re}}$ is given in Eq. \ref{eq:def_A_re}. Expressing the
heterodyne efficiency of segment 1 via Eq. \ref{eq:def_Are} and Eq.
\ref{eq:def_Bre}, results in \small
\begin{align*}
\eta_{\text{het},}^{\prime} & _{\text{seg 1}}=\\
 & \left|\int_{0}^{R_{\text{min}}}A_{\text{re}}(B_{\text{re}}+\text{i}B_{\text{im}})\,\text{d}y+\int_{0}^{R_{\text{min}}}\text{i}A_{\text{im}}(B_{\text{re}}+\text{i}B_{\text{im}})\,\text{d}y\right|^{2}\\
 & =\left|C_{1}+C_{2}\right|^{2},
\end{align*}
\normalsize with \small 
\begin{align*}
C_{1} & =\int_{0}^{R_{\text{min}}}A_{\text{re}}(B_{\text{re}}+\text{i}B_{\text{im}})\,\text{d}y,\\
C_{2} & =\int_{0}^{R_{\text{min}}}\text{i}A_{\text{im}}(B_{\text{re}}+\text{i}B_{\text{im}})\,\text{d}y,
\end{align*}
\normalsize and $C_{1},C_{2}\in\mathbb{C}.$ Moreover, we will consider
the heterodyne efficiency of segment 4 with $x^{-}=-\sqrt{R_{\text{min}}^{2}-y^{2}},\,x^{+}=0$
and $y^{-}=0,\,y^{+}=R_{\text{min}}$. Based on Eq. \ref{eq:id_error_func_3},
it is straightforward to show that 
\[
\eta_{\text{het},\text{seg 4}}^{\prime}=\left|C_{1}-C_{2}\right|^{2}.
\]
Finally, we note that $\eta_{\text{het},\text{hor}}^{\prime}$ as
given in Eq. \ref{eq:eta_hor_a_b} equals $|C_{1}|^{2}$. Using the
inner product to show that
\[
|C_{1}|\leq\text{max}\left(\left|C_{1}+C_{2}\right|,\left|C_{1}-C_{2}\right|\right),
\]
we can conclude that $\eta_{\text{het},\text{hor}}\leq\text{max}(\eta_{\text{het},\text{seg 1}},\eta_{\text{het},\text{seg 4}})$.
In a similar way it can be shown that $\eta_{\text{het},\text{ver}}\leq\text{max}(\eta_{\text{het},\text{seg 1}},\eta_{\text{het},\text{seg 4}})$.
Finally, based on symmetry arguments we have
\begin{equation}
\text{max}(\eta_{\text{het},\text{hor}},\eta_{\text{het},\text{ver}})\leq\text{max}(\eta_{\text{het},\text{seg 1}},\eta_{\text{het},\text{seg 2}}),\label{eq:max1seg_max2seg}
\end{equation}
and we can conclude that the maximum heterodyne efficiency of the
single segments is always greater or equal to the maximum heterodyne
efficiency of all possible combinations of two adjacent QPD segments.

\section{Received power\label{sec:Appendix-C-Received}}

This appendix investigates the impact of a pointing error of the
remote SC on the received power at the local SC. The modeling is performed
using the Fraunhofer approximation. This approximation is well suited
due to the large distance ($\approx2.5\cdot10^{9}$ m) between the
SC. The beam leaving the remote SC is modeled as a Gaussian beam and
is given at the external Tx pupil as follows \citep{Saleh_2019}
\begin{align}
E_{\text{s}} & (x_{\text{s}},y_{\text{s}},z_{\text{s}},t)=E_{\text{s}}^{0}\exp\left(-\dfrac{x_{\text{s}}^{2}+y_{\text{s}}^{2}}{W_{\text{tx}}^{2}(z_{\text{s}})}\right)\cdot\nonumber \\
 & \cdot\exp\left(\text{i}k_{\text{s}}z_{\text{s}}+\text{i}k_{\text{s}}\dfrac{x_{\text{s}}^{2}+y_{\text{s}}^{2}}{2R(z_{\text{s}})}-\text{i}\zeta(z_{\text{s}})-\text{i}\omega_{\text{s}}t\right),\label{eq:appC_def_tx_beam}
\end{align}
with 
\begin{align*}
W_{\text{tx}}(z) & =W_{0,\text{tx}}\sqrt{1+(z/z_{0})^{2}};\:R(z)=z\left[1+(z_{0}/z)^{2}\right];\\
\zeta(z) & =\text{atan}\left(z/z_{0}\right);\:W_{0}=\sqrt{\lambda z_{0}/\pi}.
\end{align*}
The beam waist at the Tx aperture $W_{0,\text{tx}}$ is related to
the beam waist at the QPD via $W_{0,\text{tx}}=W_{0}M_{2}$, see Fig
\ref{fig:Idenfication-of-apertures} b). The parameter $\omega_{\text{s}}$
denotes the angular frequency and $k_{\text{s}}$ the wave vector.
Finally $z_{0}$ represents the Rayleigh range. We note that the transmitting
beam, i.e. the signal beam, is expressed in its local coordinates
($x_{\text{s}},y_{\text{s}},z_{\text{s}}$). The power of an unclipped
beam is then given by
\begin{align}
P_{\text{s}'}= & \int_{A}\dfrac{1}{2}c\epsilon_{0}E_{\text{s}}\cdot E_{\text{s}}^{\ast}\,\text{d}a=\dfrac{1}{4}\pi c\epsilon_{0}W_{\text{tx}}^{2}(z_{\text{s}})\left|E_{\text{s}}^{0}\right|^{2}.\label{eq:Ptx_unclipped}
\end{align}
The electric field $E_{ff}$ in the far-field is computed in Fraunhofer
approximation. The Fraunhofer approximation at a distant $d_{\text{sc}}$,
i.e. $z_{\text{s}}=d_{\text{sc}}$, is given by \citep{Saleh_2019}
\begin{align*}
E & _{ff}(x_{\text{s}},y_{\text{s}},d_{\text{sc}},t)=h_{0}\exp\left(-\text{i}\pi(x_{\text{s}}^{2}+y_{\text{s}}^{2})/\lambda_{\text{s}}/d_{\text{sc}}\right)\cdot\\
 & \cdot\underset{A_{\text{tx}}}{\iint}E_{\text{s}}(x',y',0,t)\exp\left(\text{i}\dfrac{2\pi}{\lambda_{\text{s}}d_{\text{sc}}}(x_{\text{s}}x'+y_{\text{s}}y')\right)\text{d}x'\,\text{d}y',
\end{align*}
with $h_{0}:=(\text{i}/\lambda_{\text{s}}/d_{\text{sc}})\exp(-\text{i}k_{\text{s}}d_{\text{sc}})$,
where $\lambda_{\text{s}}$ denotes the wavelength of the signal beam.
The subscript $A_{\text{tx}}$ at the integrals indicates the integration
over the limiting aperture at the transmitting SC, which is the aperture
with radius $R_{\text{tel}}$, see Fig. \ref{fig:Idenfication-of-apertures}
b). In the following, we will express the field using cylindrical
coordinates, i.e. $E_{ff}(d_{\text{sc}},x_{\text{s}},y_{\text{s}})\rightarrow E_{ff}(d_{\text{sc}},r,\phi)$
and the integral via polar coordinates, considering the following
relation: $x_{\text{s}}=r\cos\phi,y_{\text{s}}=r\sin\phi$ and $x'=r'\cos\phi',y'=r'\sin\phi'$.
Further, using various substitutions we get: \small
\begin{align*}
E & _{ff}(r,\phi,d_{\text{sc}},t)=h_{0}\exp\left(-\text{i}\pi r^{2}/\lambda_{\text{s}}/d_{\text{sc}}\right)\cdot\\
 & \cdot\underset{A_{\text{tx}}}{\iint}E_{\text{s}}(r'\cos\phi',r'\sin\phi',0,t)e^{\text{i}2\pi\dfrac{rr'}{\lambda_{\text{s}}d_{\text{sc}}}\cos(\phi'-\phi)}\,\text{d}\phi'\,r'\text{d}r'.\\
 & =E_{\text{s}}^{0}h_{0}\exp\left(-\text{i}\pi r^{2}/\lambda_{\text{s}}/d_{\text{sc}}\right)\exp\left(-\text{i}\omega_{\text{s}}t\right)\cdot\\
 & \cdot\int_{0}^{R_{\text{tel}}}e^{-(r'/W_{0,\text{tx}})^{2}}\int_{0}^{2\pi}e^{\text{i}2\pi\dfrac{rr'}{\lambda_{\text{s}}d_{\text{sc}}}\cos(\phi'-\phi)}\,\text{d}\phi'\,r'\text{d}r'
\end{align*}
\normalsize At this point, we consider the following identity of
the zero-order Bessel function of first kind:
\begin{align*}
J_{0}(x) & =\dfrac{1}{2\pi}\int_{-\pi}^{\pi}e^{\text{i}\left(x\sin\varphi\right)}\,\text{d}\varphi=\dfrac{1}{2\pi}\int_{-(n-1)\pi+a}^{(n+1)\pi+a}e^{\text{i}\left(x\sin\varphi\right)}\,\text{d}\varphi,
\end{align*}
which is based on the periodicity of the sine. Thus, we have: \small
\begin{align*}
E & _{ff}(r,\phi,d_{\text{sc}},t)=2\pi E_{\text{s}}^{0}h_{0}\exp\left(-\text{i}\pi r^{2}/\lambda_{\text{s}}/d_{\text{sc}}\right)\cdot\\
 & \cdot\exp\left(-\text{i}\omega_{\text{s}}t\right)\int_{0}^{1}R_{\text{tel}}^{2}\rho e^{-(\rho R_{\text{tel}}/W_{0,\text{tx}})^{2}}J_{0}(\dfrac{2\pi}{\lambda_{\text{s}}}\dfrac{r}{d_{\text{sc}}}R_{\text{tel}}\rho)\,\text{d}\rho.
\end{align*}
\normalsize Note that the field is circular symmetric around the
$z_{\text{s}}$-axis and thus independent of $\phi$, i.e. $E_{ff}(r,\phi,d_{\text{sc}},t)=E_{ff}(r,d_{\text{sc}},t)$.
In the following we will express the field via the angle $\gamma_{\text{tx}}:=r/d_{\text{sc}}$
(paraxial approximation). This angle $\gamma_{\text{tx}}$ may also
be regarded as $\gamma_{\text{tx}}=\sqrt{\alpha_{\text{tx}}^{2}+\beta_{\text{tx}}^{2}}$
following the definition of section \ref{subsec:Generic-Model}. We
then get
\begin{align*}
E & _{ff}(\gamma_{\text{tx}},d_{\text{sc}},t)=\text{i}\dfrac{2\pi}{\lambda_{\text{s}}d_{\text{sc}}}E_{\text{s}}^{0}\exp(-\text{i}k_{\text{s}}d_{\text{sc}}-\dfrac{\text{i}\pi r^{2}}{\lambda_{\text{s}}d_{\text{sc}}}-\text{i}\omega_{\text{tx}}t)\cdot\\
 & \cdot\int_{0}^{1}R_{\text{tel}}^{2}\rho e^{-(\rho R_{\text{tel}}/W_{0,\text{tx}})^{2}}J_{0}(k_{\text{tx}}R_{\text{tel}}\varphi_{\text{tx}}\rho)\,\text{d}\rho\\
 & =-\dfrac{k_{\text{s}}R_{\text{tel}}^{2}}{\text{i}d_{\text{sc}}r_{\text{tx}}}E_{\text{s}}^{0}\exp(-\text{i}k_{\text{s}}d_{\text{sc}}(1+\gamma_{\text{tx}}^{2}/2-\text{i}\omega_{\text{s}}t)F(r_{\text{tx}},\kappa_{\text{tx}}),
\end{align*}
with 
\[
F(r,\kappa):=r\int_{0}^{1}\rho e^{-r^{2}\rho^{2}}J_{0}(\kappa\rho)\,\text{d}\rho
\]
and $\kappa_{\text{tx}}:=k_{\text{tx}}R_{\text{tel}}\gamma_{\text{tx}}$,
$r_{\text{tx}}:=R_{\text{tel}}/W_{0,\text{tx}}$. \\
Taking into account Eq. \ref{eq:Ptx_unclipped}, the power at the
external Rx pupil $A_{\text{rx}}$ of the local SC with radius $R_{\text{tel}}$
is given by 
\begin{align*}
P_{ff}(\gamma_{\text{tx}}) & =\dfrac{1}{2}c\epsilon_{0}\int_{A_{\text{rx}}}|E_{ff}|^{2}\,\text{d}a=\dfrac{\pi}{2}R_{\text{tel}}^{2}c\epsilon_{0}|E_{ff}|^{2}\\
 & =2P_{\text{s}'}\left(\dfrac{k_{\text{s}}R_{\text{tel}}^{2}}{d_{\text{sc}}}\right)^{2}F^{2}(r_{\text{tx}},\kappa_{\text{tx}}(\gamma_{\text{tx}})).
\end{align*}
Finally, incorporating losses in the on-board transmit and receive
path by replacing $P_{\text{s}'}$ with $P_{\text{s,n}}$ the power
of the signal beam received at the SC is given by:
\[
P_{\text{s}}(\gamma_{\text{tx}})=2P_{\text{s,n}}\left(\dfrac{k_{\text{s}}R_{\text{tel}}^{2}}{d_{\text{sc}}}\right)^{2}F^{2}(r_{\text{tx}},\kappa_{\text{tx}}(\gamma_{\text{tx}}))
\]
Note that we assumed a uniform power distribution over the integration
area at the receiving SC based on the large ratio of the distance
of the SC to the receiver aperture. This approximation is justified
by numerical evaluations, which exhibit a normalized deviation of
$1-dP_{ff}(R_{\text{tel}})/dP_{ff}(0)=5.1\cdot10^{-10}$, where $dP_{ff}$
is the received power defined over a circular area with radius $R_{\text{tel}}/1e6$.

\bibliography{ref_publication_2023_1_manuscript}

\end{document}